\documentclass[traditabstract]{aa}
\usepackage{natbib}
\usepackage{lscape}
\usepackage{wasysym}
\usepackage{adjustbox}
\usepackage{txfonts}
\usepackage{graphicx}
\usepackage{hyperref}
\bibpunct{(}{)}{;}{a}{}{,} % to follow the A&A style
\usepackage{longtable}
\usepackage{caption}
\usepackage{subcaption}
\begin{document}
 
\title{The HARPS-N Rocky Planet Search}

\subtitle{I. HD\,219134\,b: A transiting rocky planet in a multi-planet system at 6.5 pc from the Sun
\thanks{The photometric time series and radial velocities used in this work are available in electronic form at the CDS via anonymous ftp to  cdsarc.u-strasbg.fr (130.79.128.5) or via http://cdsweb.u-strasbg.fr/cgi-bin/qcat?J/A+A/} 
}

\author{
F. Motalebi\inst{1}
\and S. Udry\inst{1}
\and M. Gillon\inst{2}
\and C. Lovis\inst{1}
\and D. S\'egransan\inst{1}
\and L. A. Buchhave\inst{3,4}
\and B. O. Demory\inst{5}
\and L. Malavolta\inst{6}
\and C. D. Dressing\inst{3}
\and D. Sasselov\inst{3}
\and K. Rice\inst{7}
\and D. Charbonneau\inst{3}
\and A. Collier Cameron\inst{8}
\and D. Latham\inst{3}
\and E. Molinari\inst{9,10}
\and F. Pepe\inst{1}
\and L. Affer\inst{11}
\and A. S. Bonomo\inst{12}
\and R. Cosentino\inst{9}
\and X. Dumusque\inst{3}
\and P. Figueira\inst{13}
\and A. F. M. Fiorenzano\inst{9}
\and S. Gettel\inst{3}
\and A. Harutyunyan\inst{9}
\and R. D. Haywood\inst{8}
\and J. Johnson\inst{3}
\and E. Lopez\inst{7}
\and M. Lopez-Morales\inst{3}
\and M. Mayor\inst{1}
\and G. Micela\inst{11}
\and A. Mortier\inst{8}
\and V. Nascimbeni\inst{6}
\and D. Philips\inst{3}
\and G. Piotto\inst{6}
\and D. Pollacco\inst{14}
\and D. Queloz\inst{1,5}
\and A. Sozzetti\inst{12}
\and A. Vanderburg\inst{3}
\and C. A. Watson\inst{15}
}
%\and HARPS-N team\inst{1,3,4,5,6,7,8,9,10,11,12,13,14,...}
\institute{
Observatoire de Gen\`eve, Universit\'e de Gen\`eve, 51 ch. des Maillettes, CH-1290 Sauverny, Switzerland
\and Institut d'Astrophysique et de G\'eophysique,  Universit\'e de Li\`ege,  All\'ee du 6 Ao\^ut 17,  Bat.  B5C, 4000 Li\`ege, Belgium 
\and Harvard-Smithsonian Center for Astrophysics, 60 Garden Street, Cambridge, Massachusetts 02138, USA
\and Centre for Stars and Planet Formation, Natural History Museum of Denmark, University of Copenhagen, DK-1350 Copenhagen, Denmark
\and Cavendish Laboratory, J J Thomson Avenue, Cambridge CB3 0HE, UK
\and Dipartimento di Fisica e Astronomia "Galileo Galilei" , Universita' di Padova, Vicolo dell'Osservatorio 3, 35122 Padova, Italy
%\and INAF - Osservatorio Astronomico di Padova, Vicolo dellÕOsservatorio 5, 35122 Padova, Italy
\and SUPA, Institute for Astronomy, University of Edinburgh, Royal Observatory, Blackford Hill, Edinburgh, EH93HJ, UK
\and SUPA, School of Physics \& Astronomy, University of St. Andrews, North Haugh, St. Andrews Fife, KY16 9SS, UK
\and INAF - Fundaci\'on Galileo Galilei, Rambla Jos\'e Ana Fernandez P\'{e}rez 7, 38712 Ber\~{n}a Baja, Spain
\and INAF - IASF Milano, via Bassini 15, 20133, Milano, Italy
\and INAF - Osservatorio Astronomico di Palermo, Piazza del Parlamento 1, 90134 Palermo, Italy
\and INAF - Osservatorio Astrofisico di Torino, via Osservatorio 20, 10025 Pino Torinese, Italy
\and Instituto de Astrof\'isica e Ci\^encias do Espa\c{c}o, Universidade do Porto, CAUP, Rua das Estrelas, PT4150-762 Porto, Portugal
\and Department of Physics, University of Warwick, Gibbet Hill Road, Coventry CV4 7AL, UK
\and Astrophysics Research Centre, School of Mathematics and Physics, Queen's University Belfast, Belfast, BT7 1NN, UK
}

\date{Received on 24th of June 2015; Accepted on 30th of July 2015}

\abstract{
We know now from radial-velocity surveys and transit space missions that planets only a few times more massive than our Earth are frequent around solar-type stars.
 Fundamental questions about their formation history, physical properties, internal structure, and atmosphere composition are, however, still to be solved. We present here the detection of a system of four low-mass planets around the bright (V=5.5) and close-by (6.5 pc) star HD\,219134. This is the first result of the {\it Rocky Planet Search} program with HARPS-N on the Telescopio Nazionale Galileo in La Palma. The inner planet orbits the star in $3.0937\pm0.0004$\,days, on a quasi-circular orbit with a semi-major axis of 0.0382 $\pm 0.0003$\,AU. \textit{Spitzer} observations allowed us to detect the transit of the planet in front of the star making HD\,219134\,b the nearest known transiting planet to date. From the amplitude of the radial-velocity variation ($2.33\pm0.24$\,ms$^{-1}$) and observed depth of the transit ($359\pm38$\,ppm), the planet mass and radius are estimated to be $4.46\pm0.47$\,M$_{\oplus}$  and $1.606\pm0.086$\,R$_{\oplus}$, leading to a mean density of $5.89\pm1.17$\,g\,cm$^{-3} $, suggesting a rocky composition. One additional planet with minimum mass of $2.67\pm0.59$\,M$_{\oplus}$ moves on a close-in, quasi-circular orbit with a period of 6.765$\pm$0.005 days. The third planet in the system has a period of $46.78\pm0.16$ days and a minimum mass of $8.7\pm1.1$\,M$_{\oplus}$, at 0.234$\pm$0.002\,AU from the star. Its eccentricity is $0.32\pm0.14$. The period of this planet is close to the rotational period of the star estimated from variations of activity indicators ($42.3\pm0.1$\,days). The planetary origin of the signal is, however, the preferred solution as no indication of variation at the corresponding frequency is observed for activity-sensitive parameters. Finally, a fourth additional longer-period planet of mass of $62\pm6$\,M$_{\oplus}$ orbits the star in 1190 days, on an eccentric orbit ($e=0.27\pm0.11$) at a distance of 2.14$\pm$0.27\,AU. 
}
	

\authorrunning{Motalebi et al.}
\titlerunning{A transiting super-Earth at 6.5 pc}     
\maketitle

\section{Introduction}

The statistical analysis and occurrence rate of the small-mass planets in the range of super-Earth to Neptune are discussed in several studies with the main motivation of better understanding the orbital and physical characteristics of this population of planets, in order to constrain their formation processes \citep{Mordasini2012, Benz2014}. The radial-velocity planet search program with the HARPS spectrograph on the ESO 3.6-m telescope \citep{Mayor2003_harps, Mayor2011}, a similar survey with the Keck telescope \citep{2010Sci...330..653H}, and the NASA {\it Kepler} transit space mission \citep{2011ApJ...736...19B} have in particular contributed in a tremendous way to our knowledge of the population of small-mass/size planets around solar-type stars.

 The most common planets detected by the {\it Kepler} mission peak around 2\,R$_{\oplus}$ \citep{Howard2012, Fressin2013, Marcy2014}. Even if such planets do not exist in our Solar System, they are found around more than 30\,\% of solar-type hosts. Furthermore, many of them are found in very coplanar multiple systems \citep{2014ApJ...790..146F, Howard2012, 2012A&A...541A.139F}, tightly packed close to the central star, a new challenge to explain for planet formation models \citep{2015arXiv150403237O}. Over the last 12 years the extra-solar planet zoo has also been supplied with low-mass planets detected by the HARPS GTO planet search program and successive subsequent ESO Large programs. 
%Earlier on, and over the past 12 years, the extra-solar planet zoo was supplied as well with detections by HARPS of low-mass planets through the GTO planet search program and successive subsequent ESO Large programs. 
Notable examples include HD\,40307 \citep{ Mayor2009_SuperEarth}, HD\,10180 \citep{Lovis2011_super_earth}, HD\,20794 and HD 85512 \citep{Pepe2011_earth_likePl}, or Alpha\,Cen\,B \citep{2012Natur.491..207D}. Analysis of the results by \cite{Mayor2011} provided a list of additional super-Earth and mini-Neptunes unveiled by the survey, as well as a first statistical analysis of the properties and occurrence rate of the super-Earths and Neptune-mass planets around solar-type stars. These preliminary findings were confirmed by the fantastic statistics and precision of the {\it Kepler} detections \citep{Fressin2013}. 

After a decade focusing mainly on the detection and the determination of occurrence rate of low-mass planets, a significant observational effort is now dedicated to the planet characterization. Transit results provide the planet radius and in combination with radial velocities, the mean density of the planet can be derived. Spectral features of exoplanet atmospheres may also be revealed by space and ground-based high-resolution transmission spectroscopy in the visible and near infrared. One of the main difficulties of such characterization is the availability of bright targets favourable for follow-up observations. In particular exquisite planetary physical parameter determination will be key in lifting the intrinsic degeneracy in the determination of the planet composition when several components (gas, silicates, metals) and chemical species are mixed in the planet interior. This was a strong driver for transit search follow-up of super-Earths detected with radial velocities around very bright stars, from space with the MOST, Hubble or Spitzer space telescopes  \citep{Gillon2012_55cnce, 2010A&A...518A..25G, 2015MNRAS.450.2043D}, and for the development of a new generation of space transit missions to be launched within the coming decade: CHEOPS/ESA \citep{2014SPIE.9143E..2JF}, TESS/NASA \citep{2014SPIE.9143E..20R}, and PLATO/ESA \citep{2014ExA....38..249R}.

In order to start to fulfill the need for good estimates of planet physical parameters and considering the successful achievements with HARPS, a similar design has been implemented for the HARPS-N spectrograph. The instrument was built to become an efficient exoplanet hunter in the northern hemisphere, specifically aimed at the follow-up and characterization of the brightest transit candidates from the {\it Kepler} mission. HARPS-N is a high-precision echelle spectrograph located at the 3.6 m Telescopio Nazionale Galileo (TNG) at the Roque de los Muchachos Observatory on La Palma, Spain. It was built by an international consortium of Swiss, Italian, UK and US partners, led by Geneva university. It began operations in August 2012. Eighty nights per year for five years were granted to the consortium for a science program including the follow-up of {\it Kepler} candidates at high precision, and an additional survey, the Rocky Planet Search (RPS), to search for small planets around bright and quiet nearby stars in the northern hemisphere when \textit{Kepler} field is down.

While waiting for the new generation of ground- and space-based transit searches targeting bright stars, planet density estimates from the follow-up of the {\it Kepler} candidates with HARPS-N have already been presented in several papers \citep{Pepe2013, Dumusque2014, Bonomo2014, Dressing2015}. They are populating the low-radius regime of the mass-radius relation for small exoplanets with candidates for which precise mass and radius could be determined. 

We present in this paper the first results from the Rocky Planet Search: a planetary system around HD\,219134, composed of 3 inner super-Earths and an outer sub-Saturn mass planet, with the 4 planets spread between 0.038 and 2\,AU from the central star. Thanks to the high precision photometric observations from the \textit{Spitzer} space telescope, the inner planet is furthermore observed to transit in front of the star. After a short presentation of the Rocky Planet Search program in Sect.\,\ref{sec:RPS}, and the stellar properties of HD\,219134 in Sect.\,\ref{sec:star}, the spectroscopic and photometric observations of the new system are described in Sect. \ref{sec:planet} and \ref{sec:transit}. Sect.\,\ref{sec:discussion} provides a discussion of the stability of the system and of the composition of the inner planet. We finally draw some conclusions in Sect.\,\ref{sec:conclusion}.

\section{The HARPS-N Rocky Planet Search program} 
\label{sec:RPS}
The GTO (Guaranteed Time Observation) granted to the HARPS-N consortium is dedicated to two programs: i) the confirmation of the planetary nature and the characterization of {\it Kepler} candidates and ii) a Rocky Planet Search (RPS).
The aim of the RPS program is to perform a systematic search for low-mass planets around nearby quiet stars visible from the northern hemisphere, through an intensive monitoring of the radial velocity of quiet stars at very high precision. HARPS-N is stabilized and well-controlled in pressure and temperature to minimize instrumental drifts, and so to secure sub-m/s radial velocities over long periods of time. More details on technical aspects are presented in \cite{Cosentino2012SPIE, Cosentino2014}.

\begin{table}[h!]
\begin{center}
\begin{tabular}{lrcl}
\hline
\hline
Target ID  & Distance [pc]    & V   & Spec.Type \\
\hline
HD38          &11.75$\pm$0.38  &8.20  &K2\\
HIP1368     &14.99$\pm$0.25  &8.99  &M0\\
HD3651      &11.11$\pm$0.09  &5.88  &K0V\\
HD4628      & 7.46$\pm$0.05   &5.74  &K2V\\
HD10476    & 7.47$\pm$0.05   &5.24  &K1V\\
HD10436    &13.43$\pm$0.20  &8.42  &K5V\\
HD16160    & 7.21$\pm$0.05   &5.79  &K3V\\
HD19305    &14.77$\pm$0.39  &9.07  &M0V\\
HD232979  &10.19$\pm$0.12  &8.62  &K8V\\
HD32147    & 8.81$\pm$0.06  &6.22  &K3V\\
HIP27188   &13.69$\pm$0.21  &9.02  &M0\\
HD41593    &15.45$\pm$0.16  &6.76  &K0\\
HD47752    &17.52$\pm$0.35  &8.08  &K2\\
HD48948    &16.40$\pm$0.38  &8.59  &M0\\
HIP36357   &17.55$\pm$0.29  &7.73  &K2V\\
HD62613    &17.04$\pm$0.11  &6.55  &G8V\\
HD65277    &17.46$\pm$0.39  &8.05  &K5V\\
HD65583    &16.80$\pm$0.16  &6.97  &G8V\\
HIP39826   &17.12$\pm$0.54  &9.41  &M0\\
HIP42220   &13.90$\pm$0.24  &9.28  &M2\\
HIP43534   &16.49$\pm$0.37  &9.26  &K5\\
HD79210    & 6.19$\pm$0.20   &7.64  &M0V\\
HD79211    & 6.27$\pm$0.26   &7.70  &K2\\
HD79969    &17.53$\pm$0.23  &7.20  &K3V\\
HD84035    &17.78$\pm$0.31  &8.13  &K5V\\
HD88230    & 4.87$\pm$0.20   &6.60  &K8V\\
HIP51525   &15.67$\pm$0.27  &8.85  &K7V\\
HIP51547   &17.49$\pm$0.42  &9.63  &M0\\
HD94765    &17.55$\pm$0.26  &7.37  &K0\\
HD97101    &11.93$\pm$0.15  &8.31  &K8V\\
HD97503    &17.95$\pm$0.41  &8.70  &K5V\\
HD99492    &17.99$\pm$0.47  &7.58  &K2V\\
HD103095  & 9.16$\pm$0.07   &6.42  &G8V\\
HD110315  &14.30$\pm$0.22  &7.91  &K2\\
HD111631  &10.78$\pm$0.11  &8.49  &M0.5V\\
HD122064  &10.10$\pm$0.06  &6.49  &K3V\\
HD128165  &13.42$\pm$0.12  &7.24  &K3V\\
HD144579  &14.37$\pm$0.12  &6.66  &G8V\\
HD147379  &10.66$\pm$0.14  &8.61  &M0V\\
HD151288  & 9.77$\pm$0.08   &8.10  &K7V\\
HD158633  &12.80$\pm$0.08  &6.44  &K0V\\
HD157881  & 7.72$\pm$0.06   &7.54  &K7V\\
HD166620  &11.10$\pm$0.07  &6.38  &K2V\\
HD173818  &14.12$\pm$0.25  &8.81  &K5\\
HD185144  & 5.77$\pm$0.02   &4.67  &K0V\\
HD184489  &14.47$\pm$0.35  &9.35  &K5\\
HD190007  &13.11$\pm$0.18  &7.46  &K4V\\
HD200779  &14.84$\pm$0.30  &8.27  &K5\\
HD201091  & 3.48$\pm$0.02   &5.20  &K5V\\
HD201092  & 3.50$\pm$0.01   &6.05  &K7V\\
HD219134  & 6.53$\pm$0.03   &5.57  &K3V\\
\hline
\end{tabular}
\caption{List of the RPS targets in the GTO program on HARPS-N drawn from the Hipparcos catalogue and \cite{VanLeeuwen2007}. }
\label{tab:RPS}
\end{center}
\end{table}

{\it The sample:} The first step in establishing the RPS program was to select a sample of stars best suited for long-term precise radial-velocity measurements. Uncertainties in such observations are mainly linked to noise from three different origins: photon noise, instrumental noise, and stellar intrinsic signals. The first important criterion for this program was thus to concentrate on bright stars, in our closest neighbourhood. This is also in the interest of potential follow-up studies for characterization of the planet properties. The sample was thus selected from the Hipparcos catalogue within a distance limit set to 18\,pc\footnote{The limit was chosen iteratively in order to have a sample large enough and covering a full range of right ascension, with an over density during winter when the {\it Kepler} field in not visible}. The second step was to focus on quiet stars. Based on the CORAVEL data and investigation in the literature, we rejected known spectroscopic binaries as well as stars with $V\,\sin{i} \ge 4.5$\,kms$^{-1}$ and stars with $\log{(R^{\prime}_{HK})} \ge -4.69$ from our sample. In addition, K dwarfs are favoured in our sample compared to G dwarfs because of their lower level of stellar "noise" \citep[p-mode, granulation and activity jitter; ][]{Dumusque2011}. Their habitable zone is also closer in. Adding all the criteria together, we ended up with a sample of 51 quiet stars with a range of spectral types from G8 to M0. They are listed in Table\,\ref{tab:RPS}.

{\it Observational strategy}: To minimize the effect of stellar noise with short typical time scales (p-modes and granulations), we applied the observational strategy implemented for the HARPS very high-precision observations \citep{Pepe2011_earth_likePl}. With this strategy, we observe each target with 15 minutes exposure time to damp the p-modes effect and make a 2nd (and possibly a 3rd) measurement of the target well-spread in time during the night to damp the effect of granulation \citep[for more details see][]{Dumusque2011}. For bright stars, in order to avoid saturation the 15 minutes on target are split into sub-observations. Typical average signal-to-noise ratio (SNR) of the spectra obtained are between 200 and 400, measured at $\lambda=550$\,nm.

\begin{figure}[t!]
\begin{center}
\includegraphics[width=0.53\textwidth]{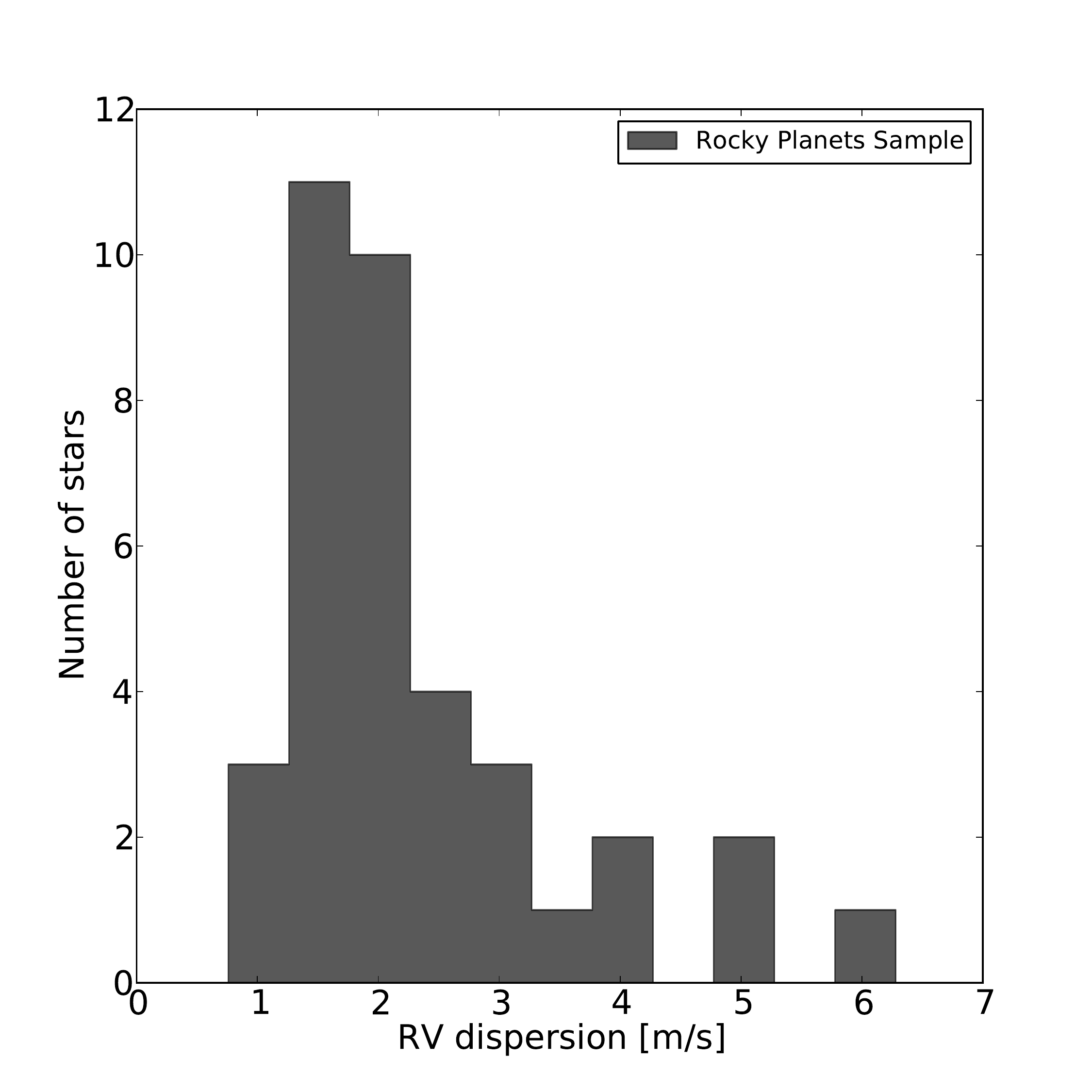}
\caption{Histogram of the dispersion of the radial-velocity measurements of stars in the RPS program on HARPS-N.}
\label{fig:histogram} 
\end{center}
\end{figure}

With continuous monitoring during the past two years of operation of HARPS-N we have acquired hundreds of precise radial velocities of the stars in the RPS sample. The typical photon-noise precision per [sub-]observation is about 0.3\,ms$^{-1}$. In Fig.\,\ref{fig:histogram}, we show the radial-velocity rms for all the stars in the RPS program. This rms naturally includes photon noise, instrumental effects (telescope, spectrograph, detector), stellar intrinsic "noise", and of course signatures of still undetected planets. With a mode around 1.48\,ms$^{-1}$, the distribution is very similar to the one obtained for the sample of solar-type stars followed at high precision with HARPS in the southern hemisphere. 

We present here the first result of the HARPS-N RPS program: the discovery of a planetary system with a transiting planet around HD\,219134. 

\section{Stellar characteristics of HD\,219134}
\label{sec:star}

The very bright star HD\,219134 has been extensively studied in the literature. Basic photometric properties of this star were taken from the Hipparcos catalogue \citep{VanLeeuwen2007}. A precise estimate of the radius of the star, $R=0.778\pm0.005$\,R$_{\odot}$, is available from interferometric measurements \citep{2012ApJ...757..112B}. Combined with a luminosity of $0.265\pm0.002$\,L$_{\odot}$, the Stefan-Boltzmann law gives $T_{\rm eff}=4699\pm16$\,K. We will adopt this value for our photometric analysis of the transiting planet in the system in Sect.\,\ref{sec:transit}. 

\begin{table}[t!]
\begin{center}
\begin{tabular}{ l c c c c}
\hline
\hline
Technique & $T_{\rm eff}$ [K] & $\log(g)$ & $\xi_{\rm t}$ & $[Fe/H]$ \\
\hline
SME$^1$            & $4835 \pm 44$ & $4.56 \pm 0.06$ &         & $0.09 \pm 0.03$ \\
LDP$^2$             & $4889 \pm 20$ & $4.60 \pm 0.20$ &         & $0.10 \pm 0.05$ \\
Photometry$^3$ & $4833 \pm 60$ & $4.59 \pm 0.02$ & 0.26 & $0.00 \pm 0.06$ \\
ULySS$^4$         & $4715 \pm 44$ & $4.57 \pm 0.07$ &         & $0.06 \pm 0.04$ \\
EWs$^5$            & $4820 \pm 61$ & $4.62 \pm 0.17$ & 0.35 & $0.12 \pm 0.04$ \\
SPC$^5$            & $4941 \pm 50$ & $4.63 \pm 0.10$ &         & $0.11 \pm 0.08$ \\
\hline
\end{tabular}
\end{center}
{\small \it $^1$\cite{Valenti2005}; $^2$\cite{2008A&A...489..923M}; $^3$\cite{Ramirez2013}; $^4$\cite{Prugniel2011}; $^5$This paper}
\caption{Comparative summary table for the atmospheric parameters derived for HD\,219134. }
\label{tab:comparison}
\end{table}

Over the past few years, the star has been the object of several studies aiming at determining photospheric parameters and chemical abundance analyses \citep{Valenti2005, 2008A&A...489..923M, Mishenina2012, Kovtyukh2003, Ramirez2013, Prugniel2011}. Results of these studies are reported in Table \ref{tab:comparison} for comparison. 

We also derived the atmospheric stellar parameters directly from HARPS-N spectra using two approaches: one based on Equivalent Width (EW) determination and one using the Stellar Parameter Classification (SPC) tool. 

For the Equivalent Width (EW) approach, we followed the procedure described in Sect.\,3 of \cite{Dumusque2014}. We used the 2014 version of the local thermodynamic equilibrium code \mbox{\sc MOOG}  \citep{Sneden1973} together with the Kurucz grid of atmosphere models \citep{Castelli2004, Kurucz1992}, while oscillator strength values in the line list from \cite{Sousa2011} were updated in accordance with the solar iron abundance $\log \epsilon ({\rm Fe \,I}) = 7.50$ from \cite{Asplund2009}.  
In order to construct high-qualitity spectra, spectra within the same visit (sub-observations) were co-added, obtaining at the time of the analysis 88 spectra with SNRs between 400 and 700. EWs were then measured for each of them and their mean and rms were used for the atmospheric parameter determination. We only retained the lines within the range $5\,m\AA < EW < 100 \,m\AA$ and with a dispersion lower than either 1\,m$\AA$ or 5\,\% of the mean EW. Despite the quality of our spectra, the preliminary analysis resulted in a poor determination of the microtubulent velocity $\xi_{\rm t}$. Following the calibration from \cite{Ramirez2013} and \cite{Tsantaki2013}, we decided to fix the microturbulent velocity to $\xi_{\rm t}=0.35$. The resulting atmospheric parameters are reported in Table \,\ref{tab:comparison}, with the gravity from FeII lines already increased by \mbox{$\Delta \log\,g$} $= 0.39\pm0.15$ according to the calibration in Sect.\,3.1 of \cite{Mortier2014}. Note that the derived parameters are dominated by systematic errors (e.g. choice of the oscillator strength, continuum placement) rather than random errors on EW measurements. 

Finally, we used the SPC tool, an alternative approach to derive atmospheric parameters by cross-correlating an observed spectrum with a library of synthetic spectra \citep{Buchhave2012Natur,Buchhave2014Natur}. With SPC we obtain an effective temperature $T_{\rm eff} = 4941\pm50$ K, a surface gravity of $4.63\pm0.10$ and $V\,\sin{i}=0.4\pm0.5$ kms$^{-1}$. The metallicity derived by SPC is $0.11\pm0.08$, from a mix of metallic absorption lines in the wavelength range between 5050 to 5360\,$\AA$. The derived values are reported in Table\,\ref{tab:comparison} as well.

The mass of HD\,219134, M$_\star=0.78\pm0.02$\,M$_\odot$, was estimated through the Synthetic Clusters Isochrones \& Stellar Tracks tool (SYCLIST)\footnote{http://obswww.unige.ch/Recherche/evoldb/index/}. SYCLIST allows the user to determine stellar parameters through a Bayesian-based interpolation of the grid of Geneva stellar evolution models. For HD\,219134, we used the $T_{\rm eff}$ taken from \cite{2012ApJ...757..112B}, the apparent magnitude and parallax from the Hipparcos catalogue \citep{VanLeeuwen2007}, and the metallicity derived from our spectral analysis as input parameters. For the latter we used the average of the values obtained from the EWs and SPC approaches.

The mean activity index log R$^{\prime}_{HK}$ is estimated from the HARPS-N spectra. We derive an average value of $-5.02$ with a dispersion of 0.06\,dex. To estimate the rotational period of the star we used a periodogram analysis of the activity indicators ($\log R^{\prime}_{HK}$, CCF FWHM and CCF bisector span time series) that yields a most significant peak at 42.3 days (see Sect.\,\ref{sec:Prot}). 

Table \ref{tab:star_summary} summarises the stellar parameters obtained from the various catalogues and analyses mentioned above, with the final values selected for the star when several estimates are available. 

\begin{table}[t!]
\begin{center}
\begin{tabular}{lccc}
\hline
\hline
Parameter &&& Ref.  \\
\hline
V                                            &                        &5.57                         &  1 \\
B-V                                         &                        &0.99                        &  1 \\
SpTyp                                    &                        &K3V                         &  1  \\
$\pi$                                      &[mas]                &$152.76\pm0.29$   & 1 \\
L$_{\star}$                            &[L$_{\odot}$]    &$0.265\pm 0.002$   &  2 \\
R$_{\star}$                            &[R$_{\odot}$]   &$0.778\pm0.005$    &  2 \\
$T_{\rm eff}$                         &[K]                   &$4699\pm16$          &  2 \\
$\log\,g$                                &[dex]                &$4.63 \pm 0.10$      & this paper $*, **$\\
$V\sin{i}$                               &[kms$^{-1}$]    &$0.4\pm0.5$           & this paper $**$\\
$\xi_{\rm t}$                          &[km\,s$^{-1}$]   &$0.35 \pm 0.19$     &  this paper $*$\\
$[{\rm Fe}/{\rm H}]$              &[dex]                 &$0.11 \pm 0.04$     & this paper $*, **$ \\
M$_{\star}$                           &[M$_{\odot}$]   &$0.78\pm0.02$       & this paper, 2 \\
<$\log\,(R^{\prime}_{HK})$>  &                        &$-5.02\pm0.06$     & this paper \\
P$_{\rm rot}$                         &[days]              &42.3 $\pm$ 0.1       & this paper\\
\hline
\end{tabular}
\caption{ Stellar parameters taken from \textit{$^1$} the Hipparcos catalogue \citep{VanLeeuwen2007}, \textit{$^2$}interferometric measurement  \citep{2012ApJ...757..112B} and estimated from HARPS-N spectra adopting \textit{$^{*}$}EWs and \textit{$^{**}$}SPC and  stellar evolution models SYCLIST for M$_{\star}$ (see text).} 
\label{tab:star_summary}
\end{center}
\end{table}

\section{Spectro-velocimetric observations}
\label{sec:planet}
\subsection{The HARPS-N data}
\begin{figure}[t!]
\begin{center}
\includegraphics[width=0.445\textwidth]{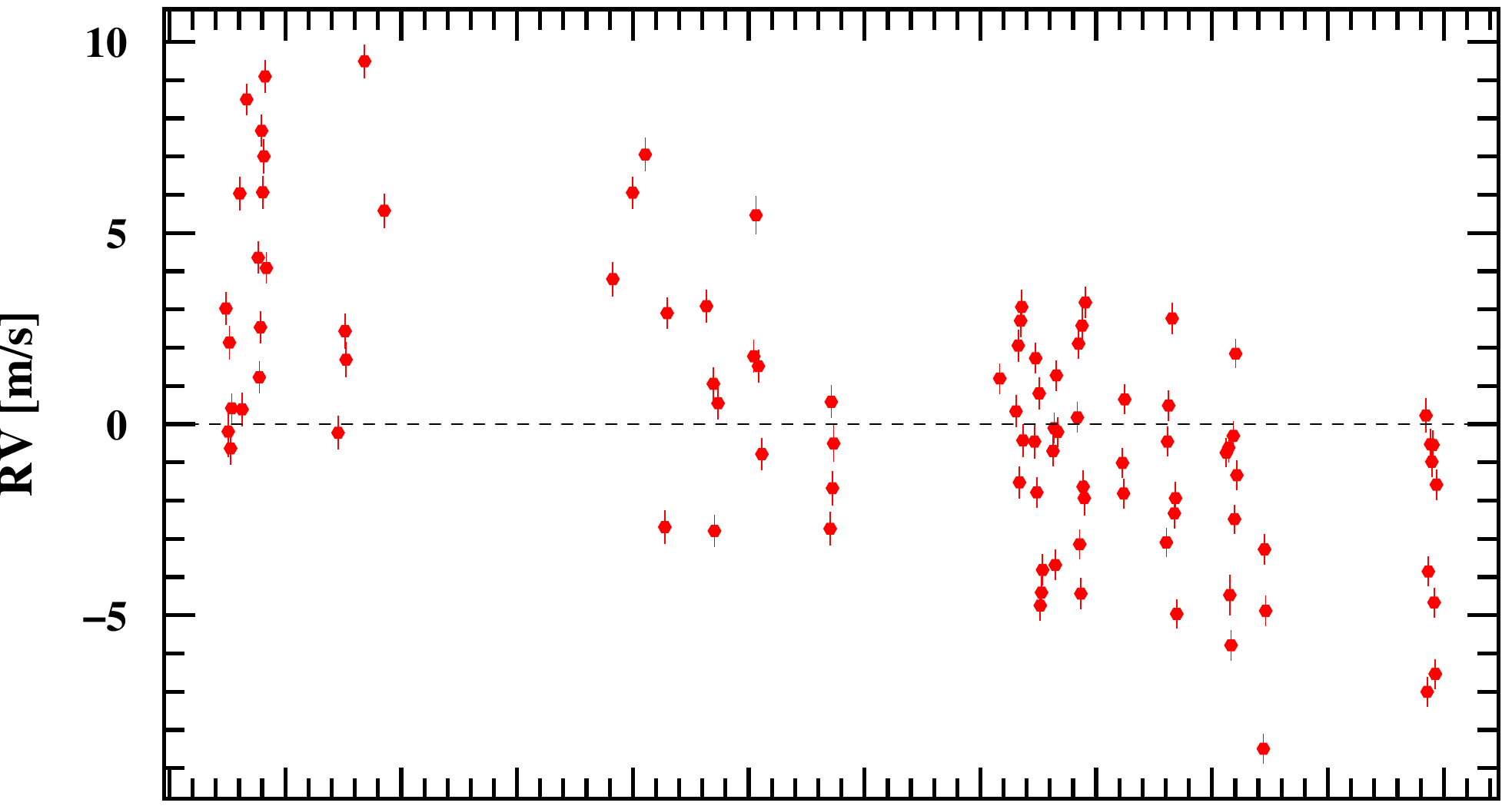}
\includegraphics[width=0.45\textwidth]{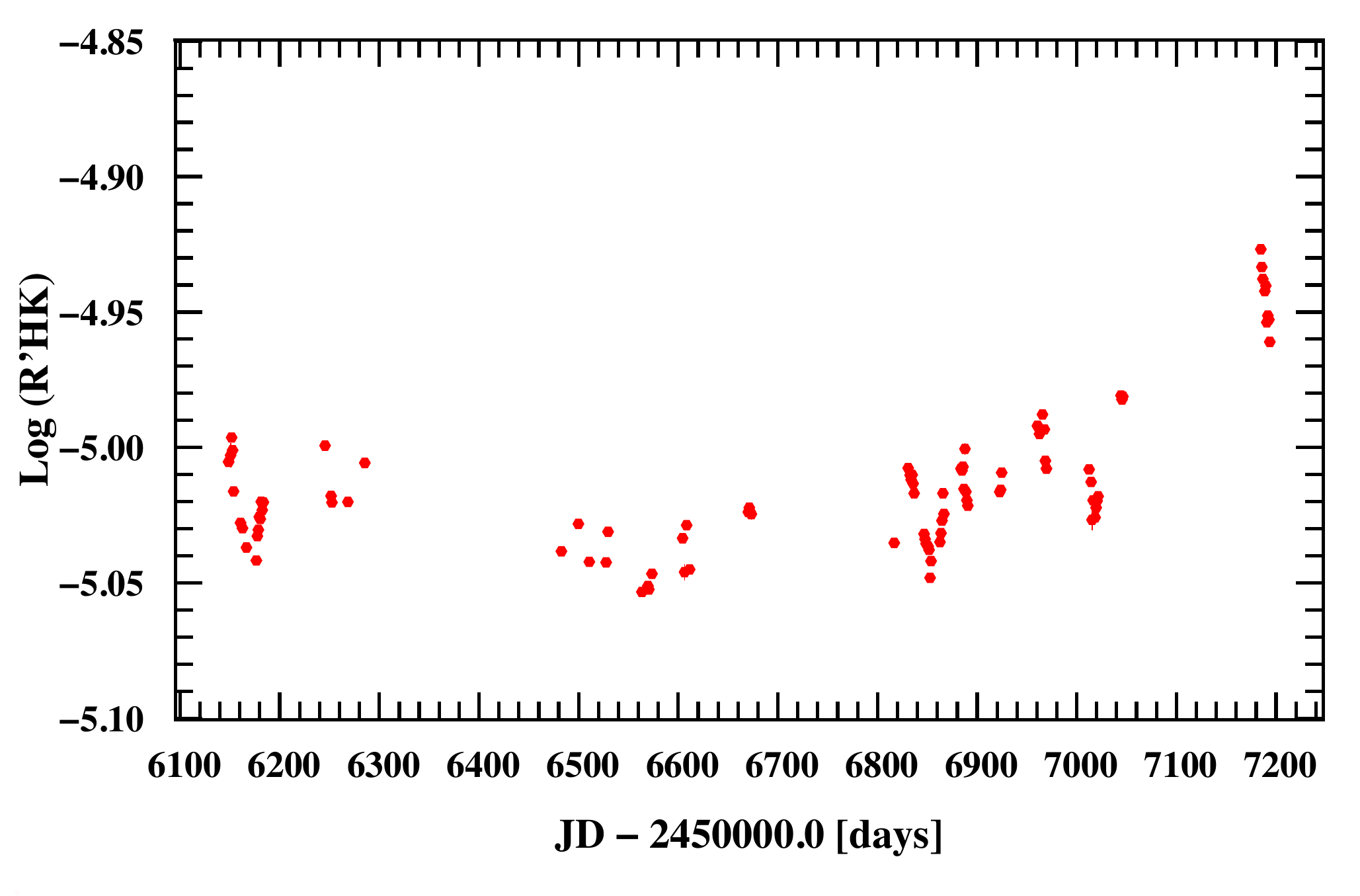}
\caption{Radial-velocity (top) and activity index (bottom) time series for HD\,219134.}
\label{fig:rv}
\end{center}
\end{figure}

HD\,219134 has been monitored by the HARPS-N spectrograph for close to three years, from BJD= 2456148.7 (9th of Aug. 2012) to BJD = 2457195.7 (22nd of June 2015). To mitigate the effects of stellar oscillations, the strategy was to stay on target for 10 to 15 minutes. Because of the brightness of the star, the observations were split in several individual sub-observations to avoid saturation. We obtained a total of 481 data points spread over 99 epochs (nightly averaged values). The nightly averaged values are displayed in Fig.\,\ref{fig:rv}. Observations were performed using the simultaneous thorium calibration technique. The mean uncertainty on the individual RVs due to photon noise and known calibration noise is 0.4 ms$^{-1}$. This corresponds to an average SNR of 268 per pixel at $\lambda$= 550\,nm. The data reduction was carried out with the latest version of the HARPS-N pipeline (DRS 3.7) using the cross-correlation technique \citep{2002A&A...388..632P, 1996A&AS..119..373B}. On top of precise radial velocities, the pipeline directly provides parameters estimated from the cross-correlation function (CCF) of the spectrum: CCF full width at half maximum (FWHM), CCF contrast (minimum vs continuum), CCF bisector span inverse slope, and Ca\,II activity index S and log R$^{\prime}_{HK}$ (Fig.\,\ref{fig:rv}).

The raw rms dispersion of the radial velocities is 3.57\,ms$^{-1}$. Once de-trended from an obvious drift, the observed radial velocities still show a dispersion of  2.77\,ms$^{-1}$, which is  significantly above the typical dispersion of such quiet stars, calling for a search for additional coherent signals in the data.

\subsection{Data analysis}
\subsubsection{Data Modelling}

The first step of the radial-velocity data analysis consists in identifying significant periodic signals in the data. This was done using the General Lomb-Scargle periodogram algorithm \citep[GLS,][]{Zechsmeister2009} applied to the nightly averaged radial-velocity measurements to which a systematic error of 1\,ms$^{-1}$ was quadratically added. False alarm probabilities were estimated  through a bootstrap approach by permuting the nightly averaged data.  Once a significant peak was located at a given period, the corresponding Keplerian was adjusted and removed. The process was repeated several times until no significant peak remained. For multiple Keplerians, all parameters were re-adjusted at each step of the analysis. During the analysis, we identified one radial velocity outlier (5\,$\sigma$ at jdb=56829.7) and decided to remove it to allow for a robust frequency analysis. 

\begin{figure}[t!]
\begin{center}
\includegraphics[width=0.45\textwidth]{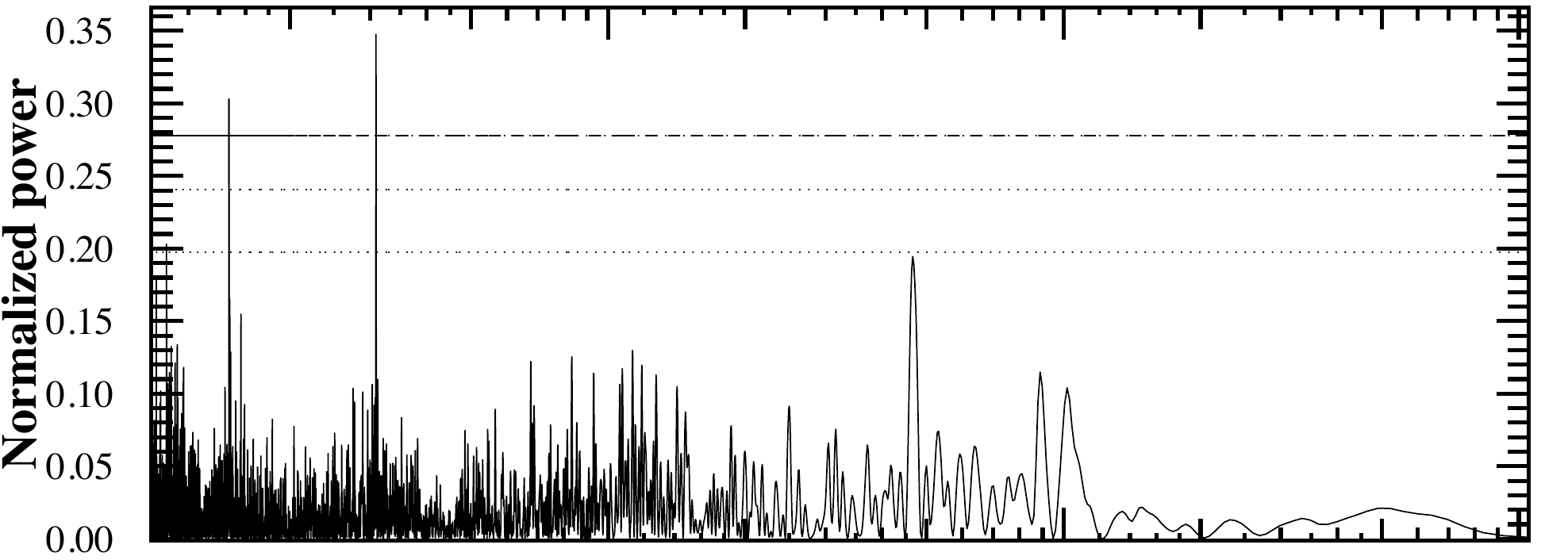}
\includegraphics[width=0.45\textwidth]{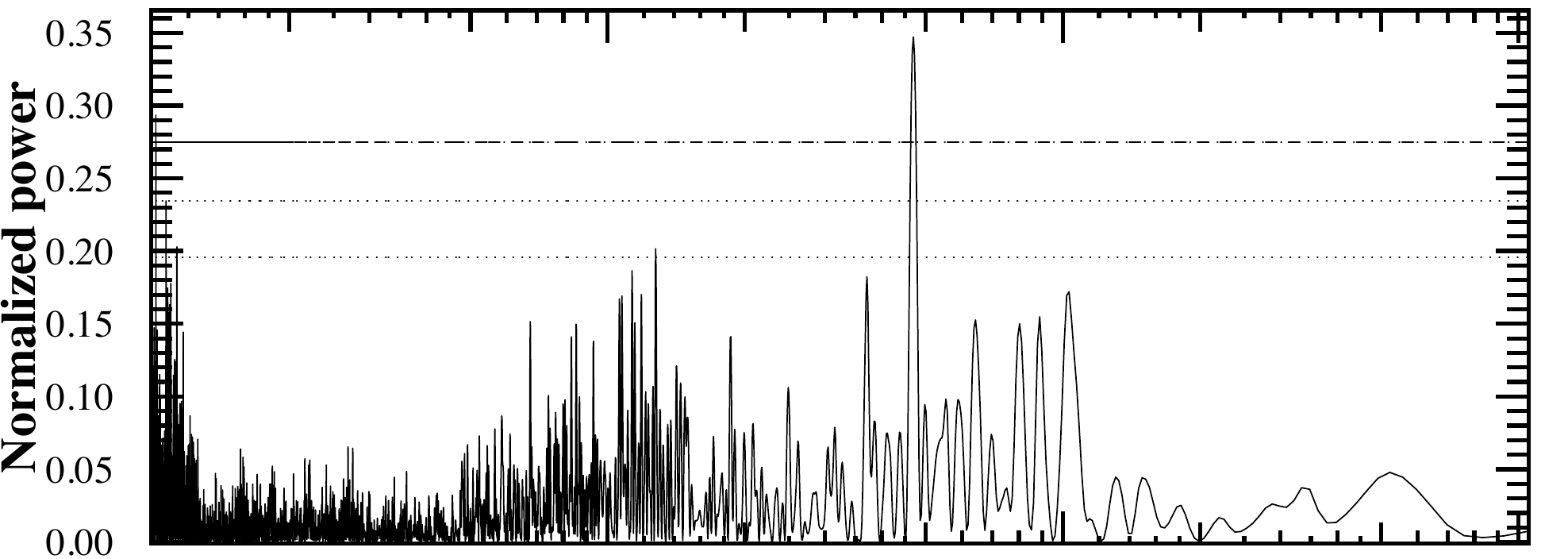}
\includegraphics[width=0.45\textwidth]{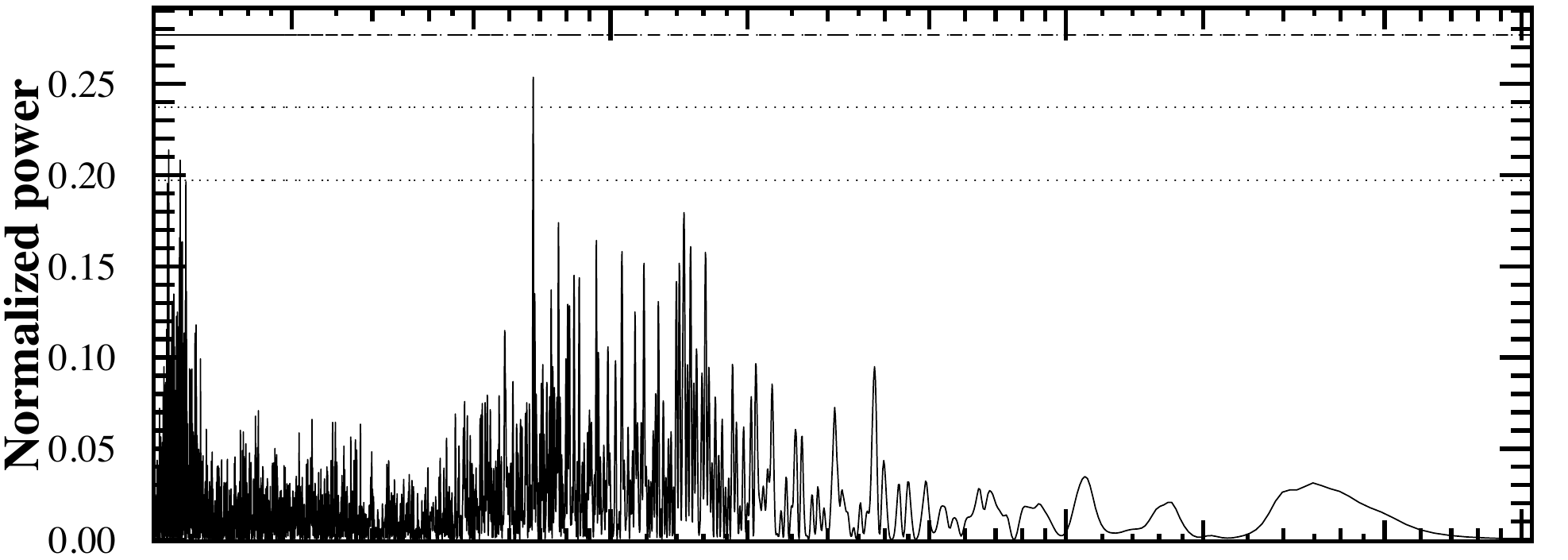}
\includegraphics[width=0.45\textwidth]{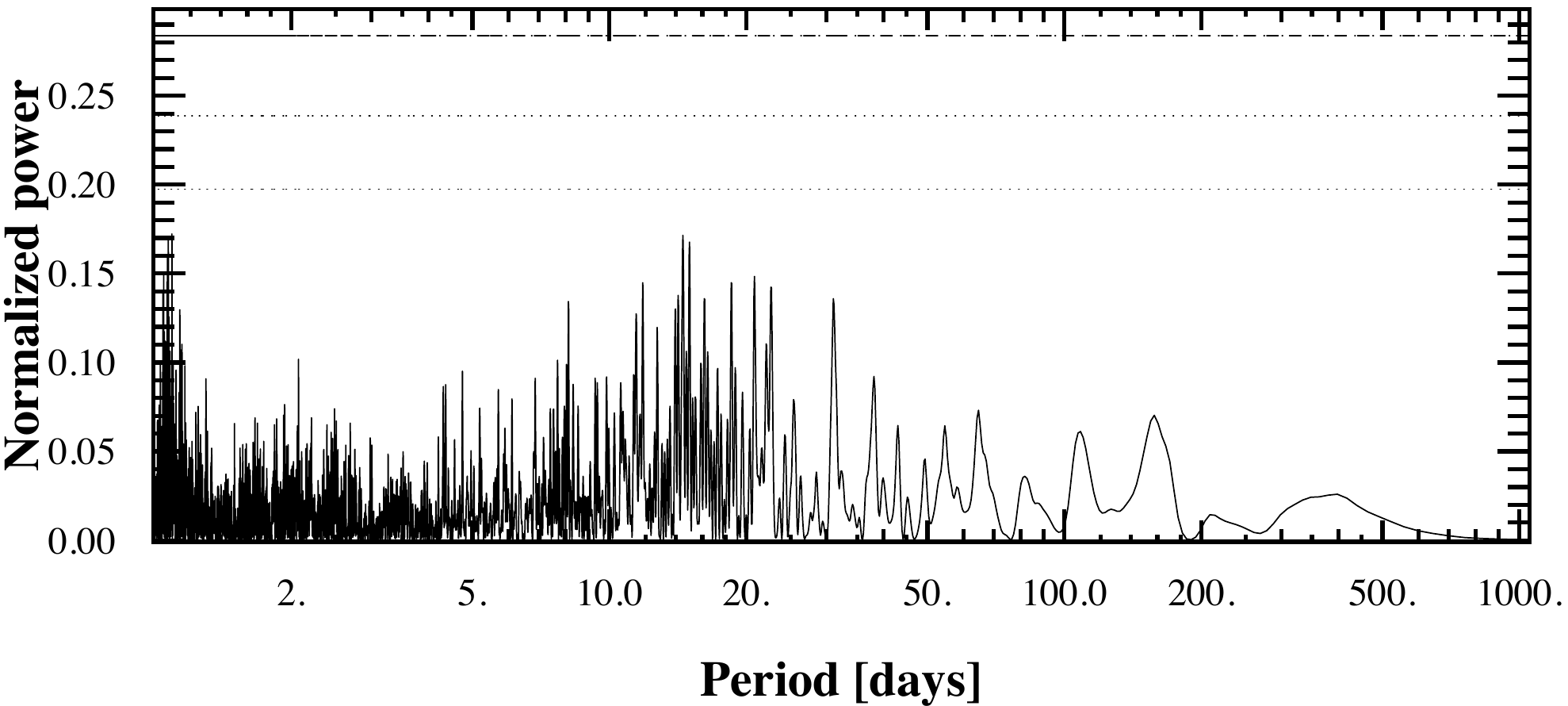}
\caption{Generalized Lomb Scargle periodogram of the radial ve- locities of HD 219134, after removing - from top to bottom - a long period Keplerian, and successively the 3.09 d signal, the 46.78 d signal and finally the 6.75 day Keplerian.  Dashed lines indicate 0.1\%, 1\%, and 10\% level of false alarm probabilities.}
\label{fig:d2_gls} 
\end{center}
\end{figure}
    
Following this procedure, a first low frequency oscillation at 1190 day was identified with a semi-amplitude of ~4.5\,ms$^{-1}$ and a 0.8\,\% false alarm probability (FAP). After removing the corresponding Keplerians, two highly significant peaks were seen in the periodogram, at 3.09 and 46.78 days, with semi-amplitudes of 2.33 and 1.94\,ms$^{-1}$, and FAPs smaller than 0.01\,\% and 1\,\%, respectively  (Fig.\,\ref{fig:d2_gls}). One additional peak remains  at 6.76 d with a significant FAP smaller than 1\,\% and a  semi-amplitude of 1.1\,ms$^{-1}$.  No more periodic signals remain in the data after subtraction of the corresponding 4 Keplerian model. This purely frequentist approach, using the nightly average data, has the advantage of being "simple and fast" and is also conservative in terms of detection limits. 

To double check the purely frequentist approach, we also analysed models of different complexity ranging from 1 to 4 Keplerian. The best model was chosen by comparing the Bayesian Information Criterion \citep[BIC,][]{Kass1995} between the different models:

$$
{\rm BIC} = -2 \log \L + N_{\rm param} \log N_{\rm meas}  \quad, 
$$
where $\log \L $ is the maximum of the log likelihood, $N_{\rm param}$ the number of free parameters in the model, and $N_{\rm meas}$ is the number of measurements. In model comparisons based on the Bayes factor, when the prior information is smaller than the information provided by the data, the BIC (a.k.a Schwarz criterion) indicates that the model with the highest probability is the one with the minimum BIC. A summary of the comparison between a few models is provided in Table\,\ref{tab:models}. The  4 Keplerian model has the lowest BIC and differs from the 3 Keplerian model by $\Delta$BIC=34. It is usually assumed that a   $\Delta$BIC of 20  between two models is considered as strong evidence in favour of the most complex one which leads us to adopt the 4 Keplerians as our best solution. \\

\begin{table}[t!]
\begin{center}
\begin{tabular}{lcccc}
\hline
\hline
Model (Period [days])&BIC&$\Delta$BIC&$\chi^{2}_{r}$& \multicolumn{1}{p{1cm}}{\centering $\sigma_{o-c}$ \\ $[ms^{-1}]$}\\
\hline
K1 (1190)&611& ~    &6.35 &2.63\\
K2 (1190, 3.09)& 389 &221 &3.90 &2.00\\ 
K3 (1190, 3.09, 46.8)&275&114 &2.47 &1.54 \\
K4 (1190, 3.09, 46.8, 6.76)&241 & 34  &1.89 &1.31   \\
\hline
\end{tabular}
\caption{Models tested on the 98 nightly averaged radial velocities with an additional noise of 1\,ms$^{-1}$ quadratically added to the radial-velocity error bars. The 4 Keplerian model is preferred since it has the lowest  BIC, $\chi^{2}_{r}$ and residual dispersion. These values are estimated based on the maximum likelihood. }\label{tab:models}
\end{center}
\end{table}

To obtain robust confidence intervals for the Keplerian parameters as well as an estimate of the additional noise present in the data (nuisance parameter below), we further probe the parameter space with a Markov Chain Monte Carlo algorithm (MCMC) with Metropolis-Hasting. An abundant literature discusses in much detail the implementation of MCMC posterior sampling (\cite{Andrieu:2008kh} for a pure statistical approach; \cite{2005ApJ...631.1198G, 2005blda.book.....G},  \cite{2007MNRAS.380.1230C}, and \cite{2008MNRAS.385.1576P}  for exoplanet searches). Our MCMC probes the following set of parameters: $\log{p}$, $\sqrt{e}\cos{\omega}, \sqrt{e}\sin{\omega}$,  $\log{K}$, $\lambda_{0}$ (the mean longitude at a given epoch) while the noise model follows a simple normal law with standard deviation derived from the observation errors and a nuisance parameter ($\sigma_{\rm tot}^{2}=\sigma_i^{2} + s^{2}$). Jeffrey's priors are used for the period, the radial-velocity semi-amplitude, and the nuisance parameter while uniform priors are used for the other parameters. 

\begin{figure}[t!]
\begin{center}
\includegraphics[width=0.5\textwidth]{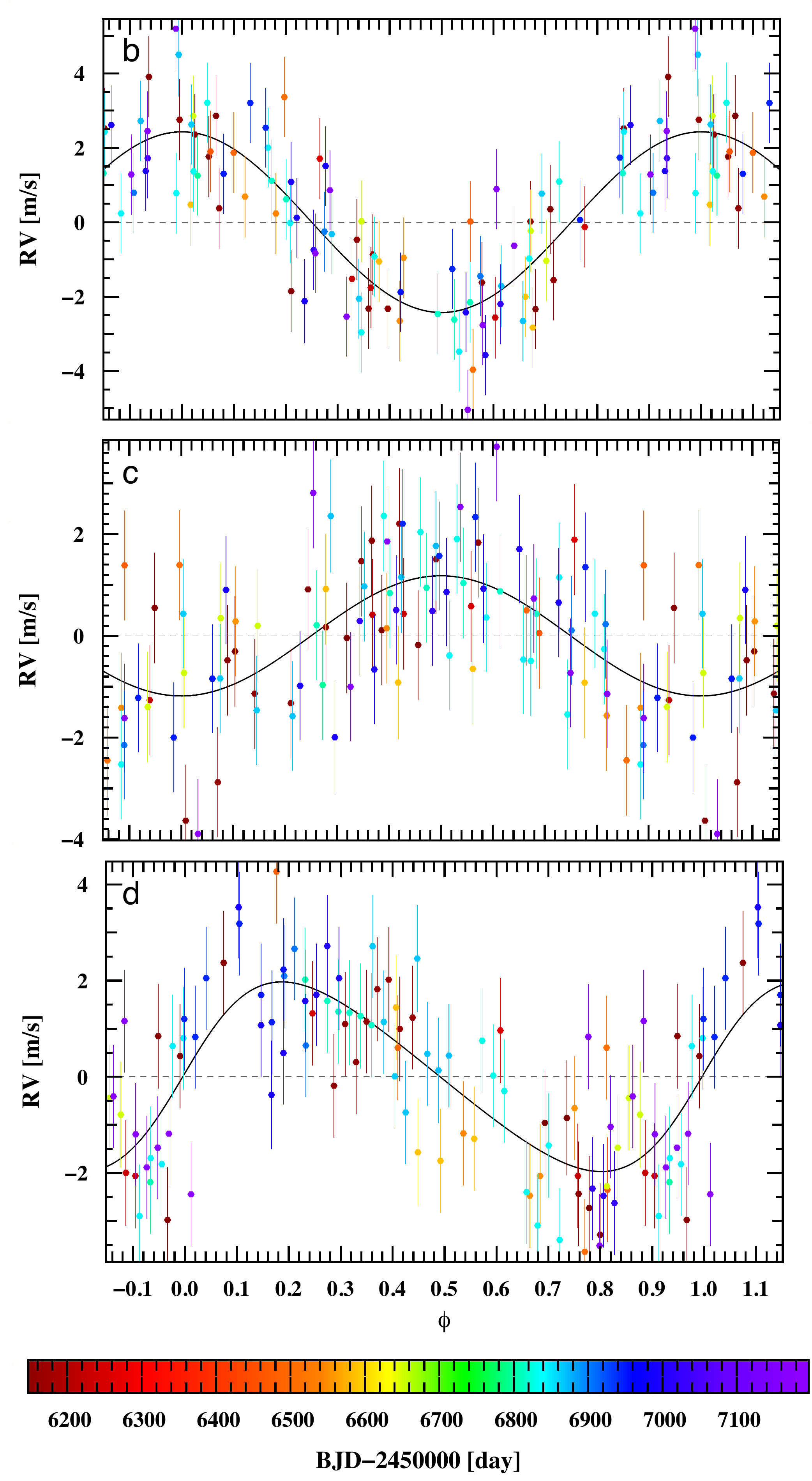}
\caption{Phase-folded radial-velocity measurements of HD\,219134 with the corresponding Keplerian model (solid line) for each of the 3 inner super-Earths, after removing the contribution of all the other planets in the system. From top to bottom, we have the 3.09-, 6.76-, and 46.78-day periods.}
\label{fig:pl1} 
\end{center}
\end{figure}

\subsubsection{Orbital solution }
\begin{figure}[t!]
\begin{center}
\includegraphics[width=0.45\textwidth]{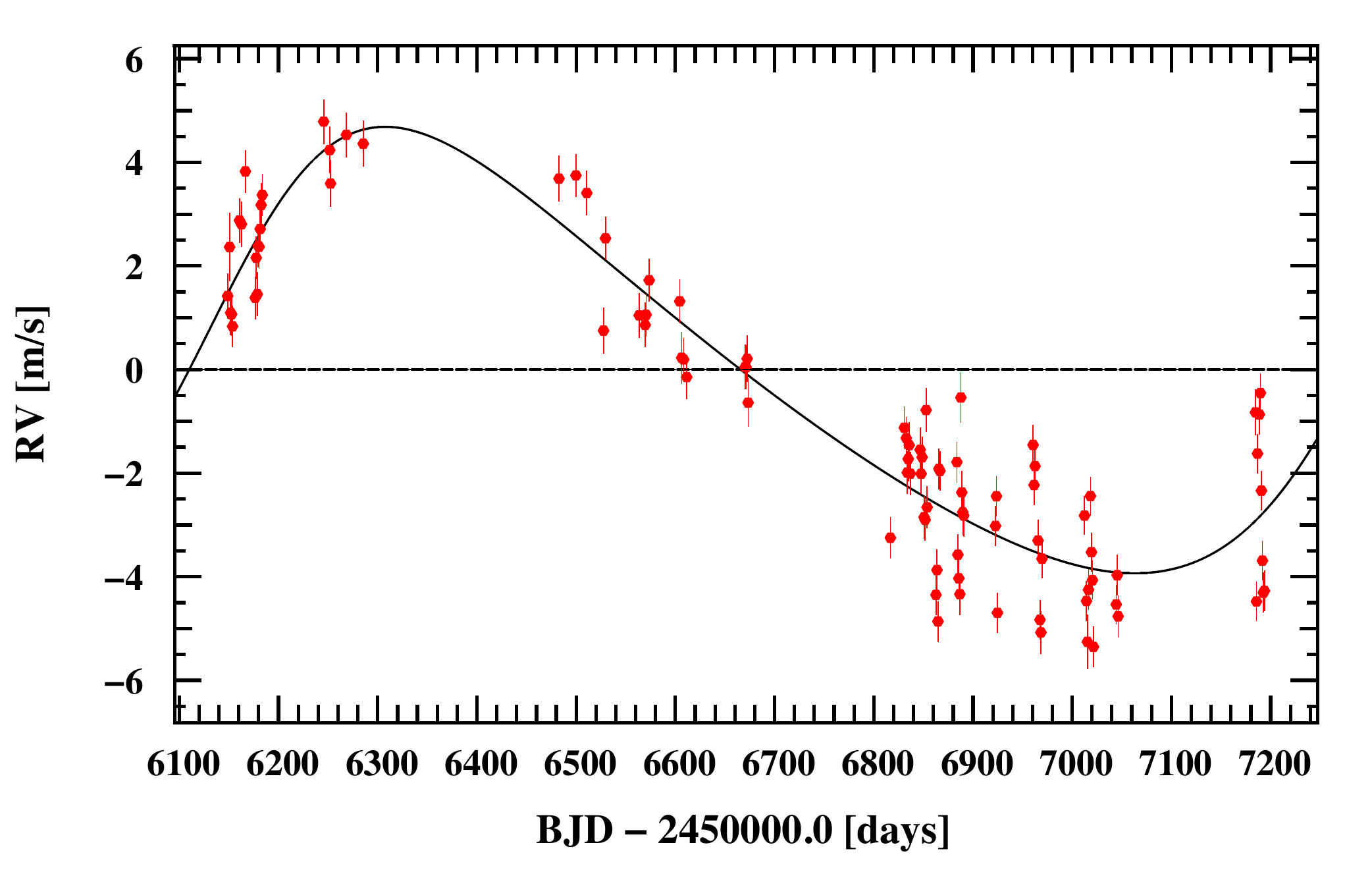}
\caption{Radial-velocity measurements as a function of time of HD\,219134 with the corresponding Keplerian model (solid line) for the outer planet in the system, after removing the contribution of the 3 inner ones.}
\label{fig:pl5} 
\end{center}
\end{figure}

The orbital elements corresponding to the four Keplerian model and transit timing constraint are listed in table~\ref{tab:planet-solution} while the phase folded radial velocities are displayed in figures~\ref{fig:pl1} and \ref{fig:pl5}.
The shortest period signal, with a radial velocity semi-amplitude of 2.33\,ms$^{-1}$ at 3.09 days, is clearly identified as a super-Earth (4.46\,M$_{\oplus}$) in a circular orbit. One additional low amplitude signal (K=1.1\,ms$^{-1}$) is present at 6.76 days, corresponding to a planet with minimum mass of 2.67\,M$_{\oplus}$. The existence of this signal is corroborated by both the FAP and the BIC estimators but its strength might be affected by the un-perfect modelling of the long-period signal (see below).

 The third signal with a period of 46.78 days and a semi-amplitude of 1.94\,ms$^{-1}$  corresponds to a super-Earth of 8.67 \,M$_{\oplus}$ with an eccentric orbit of $e=0.32$.  We show, in the following section, that this period, which is close to the  rotational period of the star (42.3 days), is not of stellar origin. Its relatively high eccentricity may also originate from an un-perfect modelling of the outer signal.

 The long period radial velocity trend clearly shows two extrema, which favour a Keplerian  instead of a polynomial drift, with a period of  1190 days and a semi-amplitude of 4.5 ms$^{-1}$. As explained below, its origin cannot be linked to any long term stellar activity fluctuation and is undoubtedly of planetary origin. Due to the incomplete coverage of the orbit, both the eccentricity and the period of the planet remain moderately constrained but still leads to a good estimate of its minimum-mass, {\it i.e.} 62\,M$_{\oplus}$.

\begin{table*}[t!]
\begin{center}
\begin{tabular}{llccccc}
\hline
\hline
Model          &                &\multicolumn{4}{c}{K4+ $\mathcal{N}\Big(0,\sqrt{\sigma_{i}^{2}+s^{2}}\Big)$}             \\
                    &                & HD\,219134\,b         &HD\,219134\,c            &HD\,219134\,d &HD\,219134\,e\\
\hline
$P$              & [days]   &$3.0937\pm0.0004$  &$6.765\pm0.005$             &$46.78\pm0.16$& $1190^{+379}_{-34}$\\
$K$              & [m/s]     &$2.33\pm0.24$         &$1.09  \pm0.26$              &$1.94\pm0.29$  &$4.46\pm0.52$ \\
$\lambda_{0}$\, & [deg]    &$82\pm8$           &$295\pm20$        &$98\pm16$      &$206^{+3}_{-45}$ \\
$T_{t}$       & [BJD-2400000]   &$57126.7001\pm0.001$  &$57129.46\pm0.45$   &~ &~\\
$\sqrt{e}.\cos({\omega})$ &      &$0.05\pm0.19$           &$0.17\pm0.26$                 &$-0.43\pm0.18$&$0.21\pm0.23$ \\
$\sqrt{e}.\sin({\omega})$  &      &$-0.11\pm0.21$          &$-0.03\pm0.31$                &$0.03\pm0.21$ &$-0.35\pm0.24$\\

\hline
$e$               &                 &$0.00^{+0.13}_{-0.00}$         &$0.00^{+0.26}_{-0.00}$         &$0.32\pm0.14$ &$0.27\pm0.11$ \\
$\omega$    &                & undefined  & undefined    &$143\pm33$&$288\pm45$ \\
$m_{pl}\sin{i}$  &[M$_{\oplus}$] &$4.46\pm0.47$      &$2.67\pm0.59$            &$8.67\pm1.14$   &$62\pm6$ \\
$a$                   & [AU]                &$0.0382\pm0.0003$  &$0.064\pm0.001$   &$0.234\pm0.002$ &$2.14^{+0.43}_{-0.02}$ \\
\hline
$\gamma$   &                         &\multicolumn{4}{c}{$-18.4203$\, kms$^{-1}$ $\pm0.6$\, ms$^{-1}$ }   \\
%\hline
$N_{meas}$     &                             &\multicolumn{4}{c}{98}             \\
s                      &        [m/s]      &\multicolumn{4}{c}{$1.18 \pm 0.06$ }             \\
\hline 
\end{tabular}
\caption{Orbital solution for the 4 Keplerian model (K4) and planet inferred parameters for the system around HD\,219134. $T_{t}$ is the expected date of transit. $\lambda_{0}$ is the mean longitude at the time 2457126.7001 day, corresponding to the transit timing (see Sect.\,\ref{sec:transit})}
\label{tab:planet-solution}

\end{center}
\end{table*}

\begin{figure}[h!]
\begin{center}
\includegraphics[width=0.45\textwidth]{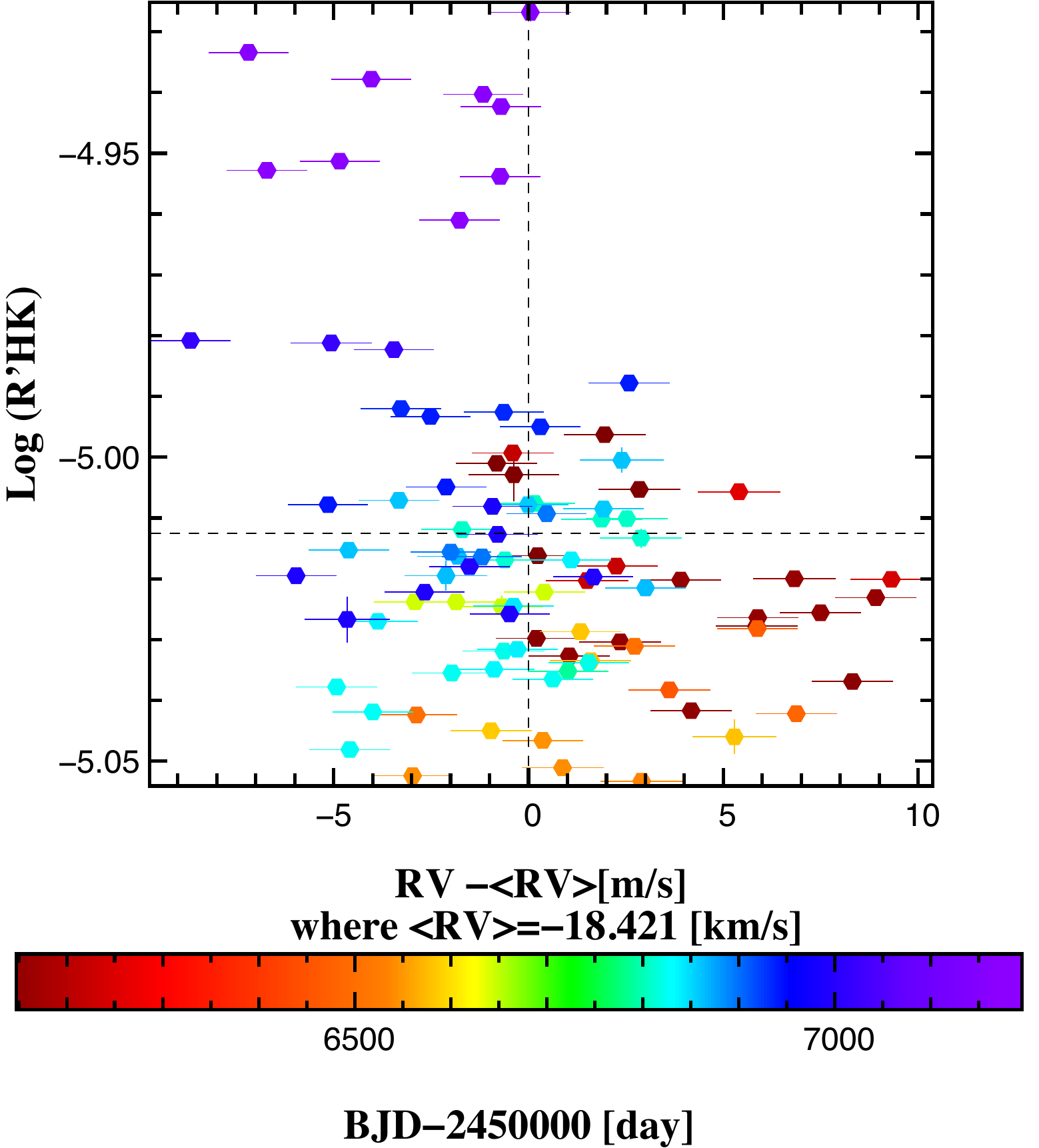}
\caption{Plot of the star activity index as a function of the radial-velocity residuals after removing the 3 shorter-period planet contributions for HD\,219134. No clear positive correlation is observed, contrary to what is expected in the case of magnetic cycles \citep{Lovis2011_magnetic}.}
\label{fig:rv-rhk}
\end{center}
\end{figure}

\subsubsection{Periodic signals: planet vs activity-related origin }
\label{sec:Prot}

{\it Long-period magnetic cycle:} The radial-velocity and activity index $\log\,R^{\prime}_{HK}$ measurements of the star HD\,219134 are displayed in Fig.\,\ref{fig:rv}. Although both exhibit long-term variations, the $\log\,R^{\prime}_{HK}$ time series is very stable over the first 750 days of the survey and slightly increases over the last 300 days whereas the radial velocity constantly decrease over the ~1050 day observation time span. No correlation is observed between RVs and log R$^{\prime}_{HK}$  as shown on Fig.\,\ref{fig:rv-rhk}. The long term trend of the $\log\,R^{\prime}_{HK}$ is most probably related to the magnetic cycle of the star  \citep{Lovis2011_magnetic, 2011A&A...535A..55D, 2013A&A...551A.101M} while the observed 1190~day period  corresponds to a long period companion.  Moreover, at the observed level of activity, we expect a low impact of stellar activity on the observed radial velocities.

\begin{figure}
\begin{center}
\includegraphics[width=0.47\textwidth]{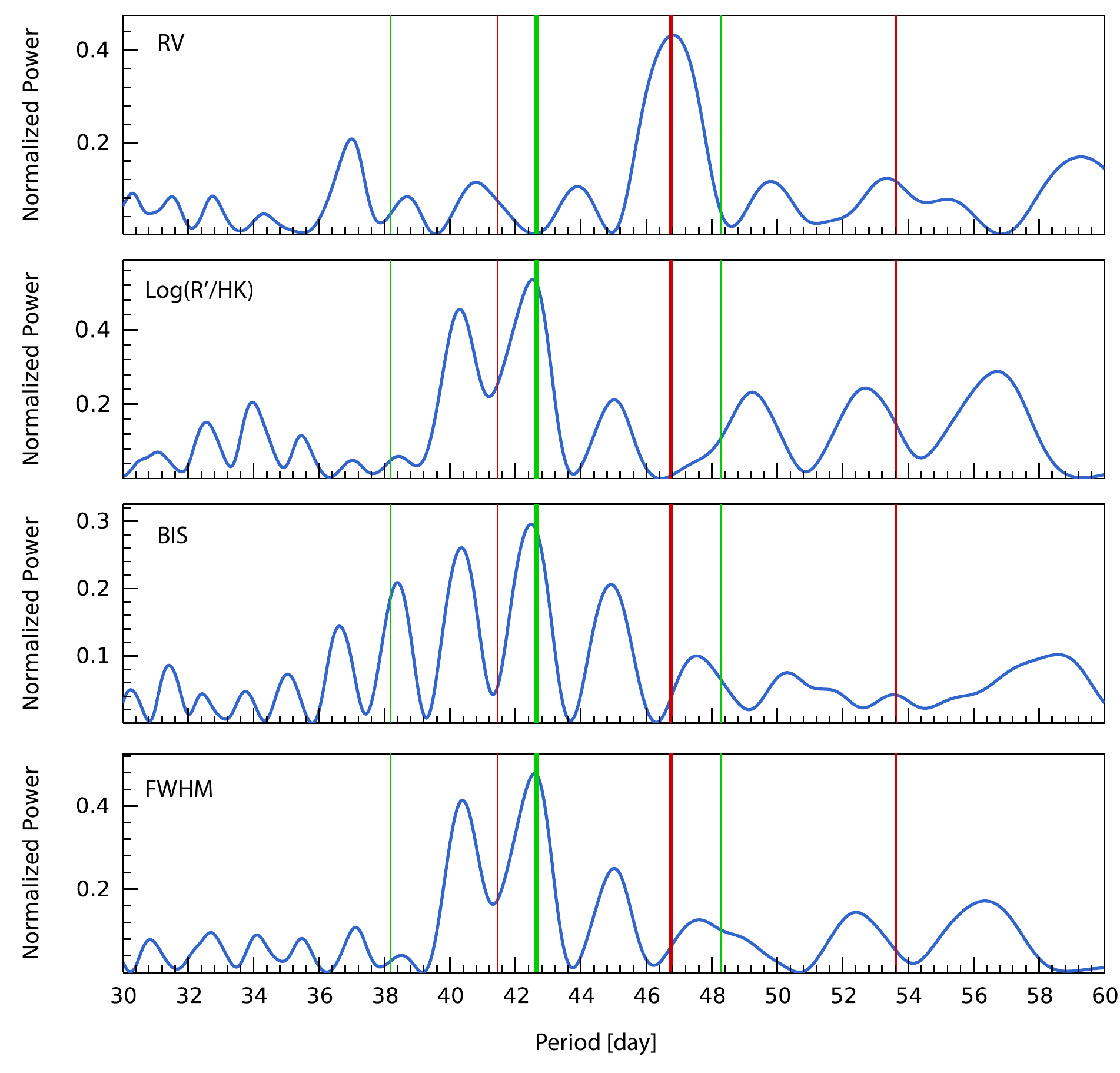}
\caption{From {\it Top} to {\it Bottom} the panels show the GLS periodograms of radial velocities, $\log\,R^{\prime}_{HK}$, CCF bisector span, and CCF FWHM of the 2-year data set, zoomed around the periods of interest. The thick red line in each panel, indicates the planetary period (46.8 days) and the thick green line indicates the stellar rotation period (42.3 days). The thinner lines are the corresponding yearly aliases.}
\label{fig:alias}
\end{center}
\end{figure}

{\it Rotation period:} Although different, the period of the third planet (46.8 days) is not very far from the rotational period of the star estimated from coherent variations of the activity indicators. A periodic variation around 42.3 days is indeed observed in the log R$^{\prime}_{HK}$ as well as in the CCF FWHM and in the bisector span of the radial velocities, whereas no signal appears around 46.8 days, as illustrated in Fig.\,\ref{fig:alias} showing the GLS periodograms of these parameters zoomed around the periods of interest. We can therefore safely consider the 42.3 days as a valid estimation of the rotational period of the star (P$_{rot}$). Conversely, no signal at 42.3 days appears in the radial-velocity data, supporting the planetary solution.

In order to avoid any misinterpretation of the stellar activity as a planetary signal, we pushed our investigations a bit further. We first examined the yearly aliases \citep{Dawson2010} of the 46.8 and 42.3 days signals (Fig.\,\ref{fig:alias}) and confirmed that one period is not an alias of the other. In a second step, we considered several subsets of the data to check the persistence of the planetary signal over time and to mitigate the possible effects of discontinuities in the data sampling. 
The same features are observed. This confirms that the planetary signal at 46.8 days is present in the radial velocities at any time, and that in the same data no signal appears at the estimated $P_{rot}$ (42.3 days). 

\begin{figure}[t!]
\begin{center}
\includegraphics[width=0.5\textwidth]{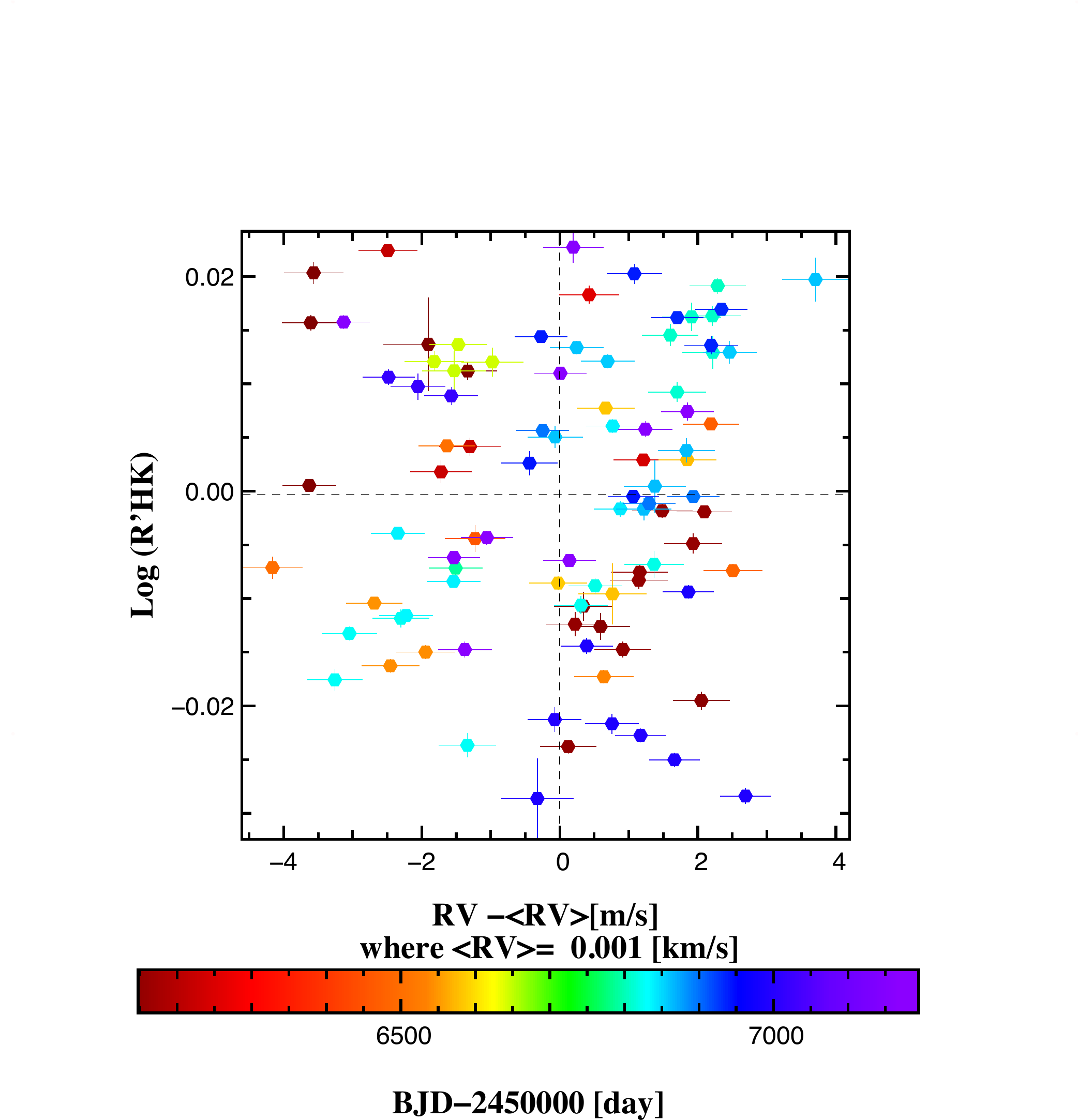}
\caption{Radial velocities after subtracting the contributions of the planets at 3.09, 6.76 and 1190 days, displayed as a function of the $\log R^{\prime}_{HK}$ activity index detrended with a polynomial of degree 3. No correlation is observed supporting so the planetary interpretation of the 46.8-day signal still present in the data.  }
\label{fig:rv-rhk-46}
\end{center}
\end{figure}

A final argument in favour of the planetary interpretation of the 46.8 day signal is provided in Fig.\,\ref{fig:rv-rhk-46} by the absence of a correlation between the $\log R^{\prime}_{HK}$ activity index and the residuals around a 3-planet solution (leaving the 46.8 day period out). Such a correlation would be expected if the radial-velocity variation is induced by activity-related spots or plages on the star surface.

\section{Transit detection with \textit{Spitzer} space telescope}
\label{sec:transit}
\subsection{Spitzer observations}

Thanks to its short-period of 3.09\,d, HD\,219134\,b has an interestingly high geometric  transit probability of $\sim$9\%. In Feb 2015, we requested 9.5hr of Director's Discretionary Time (DDT) on the {\it Spitzer Space Telescope} to continuously monitor the 2-$\sigma$ transit window of the planet as derived from our analysis of the HARPS-N RVs. As demonstrated by its co-detection and subsequent studies of the transits and occultations of 55\,Cnc\,e \citep{2011A&A...533A.114D, 2012ApJ...751L..28D, 2015arXiv150500269D, Gillon2012_55cnce} and its confirmation of the transiting nature of HD\,97658\,b \citep{Van_Grootel2014}, {\it Spitzer} is indeed a very powerful facility to search for and measure with high-precision the transits of low-mass RV planets, thanks to its trailing orbit that allows monitoring the same star continuously for a complete transit window, and to its ultra-high photometric precision (a few dozens of ppm per time interval of 30 min for 55\,Cnc).  Our program was approved by the {\it Spitzer Science Center} (SSC) Director under the ID\,11180 (PI: M. Gillon), and the SSC managed to schedule it for 2015 Apr 14th, a few days before the end of the current visibility window of the star. 

We observed HD\,219134 at 4.5 $\mu$m with the {\it Spitzer}/IRAC detector \citep{2004ApJS..154...10F} in subarray mode (32x32 pixels windowing of the detector), the extremely fast Fowler sampling ($\sim$0.01s)  maximizing the duty cycle and SNR. No dithering pattern was applied to the telescope (continuous staring). HD\,219434  being an extremely bright star for {\it Spitzer}, we used the shortest available integration time of 0.01s, allowing the counts to remain in the linear regime of the detector. The observations were performed from 2015 Apr 14th 01h58 to 11h18 UT. We used the  recently introduced PCRS peak-up mode \citep{Grillmair2012, Ingalls2012} which was previously used by \cite{2014ApJ...790...12B} to estimate the infrared transit depth of Kepler-93\,b. This mode provides enhanced accuracy in the position of the target on the detector, leading to a significant decrease of the so-called `pixel phase effect' that is the most important source of correlated noise in high-SNR staring mode observation with IRAC InSb arrays \citep[e.g. ][]{2008ApJ...673..526K}. The run consisted of a 9hr-long science Astronomical Observational Requests (AOR) preceded by a short (30 min)  AOR to allow the spacecraft to stabilize.

\subsection{Data reduction}

After download to Earth and basic calibration with the {\it Spitzer} pipeline S19.1.0, the images were made available to us by SSC through the Spitzer Heritage Archive (SHA) web interface\footnote{http://sha.ipac.caltech.edu} under the form of Basic Calibrated Data (BCD). Each subarray mode BCD is composed of a cube of 64 subarray images of 32$\times$32 pixels (pixel scale = 1.2 arc second). We used the following strategy to reduce these BCDs. We first converted fluxes from the {\it Spitzer} units of specific intensity (MJy/sr) to photon counts, then aperture photometry was performed on each subarray image with the {\tt IRAF/DAOPHOT}\footnote{IRAF is distributed by the National Optical Astronomy Observatory, which is operated by the Association of Universities for Research in Astronomy, Inc., under cooperative agreement with the National Science Foundation.} software \citep{1987PASP...99..191S}. We tested different aperture radii, and selected 2.3 pixels as the radius minimizing at best the white and red noises in the residuals of a short data fitting analysis. The centre and width of the Point-Spread Functions (PSF) were measured by fitting a 2D-Gaussian profile on each image. The $x-y$ distribution of the measurements was then looked at, and measurements having a visually discrepant position relative to the bulk of the data were then discarded. For each block of 64 subarray images, we then discarded the discrepant values for the measurements of flux, background, $x$- and $y$-positions using a 10-$\sigma$ median clipping for the four parameters, and the resulting values were averaged, the photometric errors being taken as the errors on the average flux measurements. Finally, a 50-$\sigma$ slipping median clipping was used on the resulting light curves to discard outliers (due, e.g., to cosmic hits). 

Our resulting light curve counted 9396 measurements. It is shown in Fig. \ref{fig:S1}, with the evolution of relevant external parameters (PSF $x$- and $y$-center and  PSF width, background). Its time sampling $\sim3.4s$ being much shorter than the structures of the expected transit and of the typical timescale of the {\it Spitzer} systematics, we binned the light curve to time intervals of 30s for the sake of computational speed of the data analysis. 
Nevertheless, we verified with a shorter version of the data analysis procedure described below that our results are insensitive to the binning of the photometry.

\subsection{Data analysis}

We analysed the {\it Spitzer} photometric time-series with our adaptative MCMC code (see Gillon et al. 2012 and references therein). The assumed photometric model consisted of the eclipse model of \cite{2002ApJ...580L.171M} to represent the possible transit of HD\,219134\,b, multiplied  by a baseline model aiming to represent the other astrophysical and instrumental effects at the source of photometric variations. We assumed a quadratic limb-darkening law for the star. We based  the selection of the baseline model on the minimization of the Bayesian Information Criterion \citep[BIC,][]{Schwarz1978}. 

\begin{figure}[t!]
\centering                     
\includegraphics[width=0.5167\textwidth]{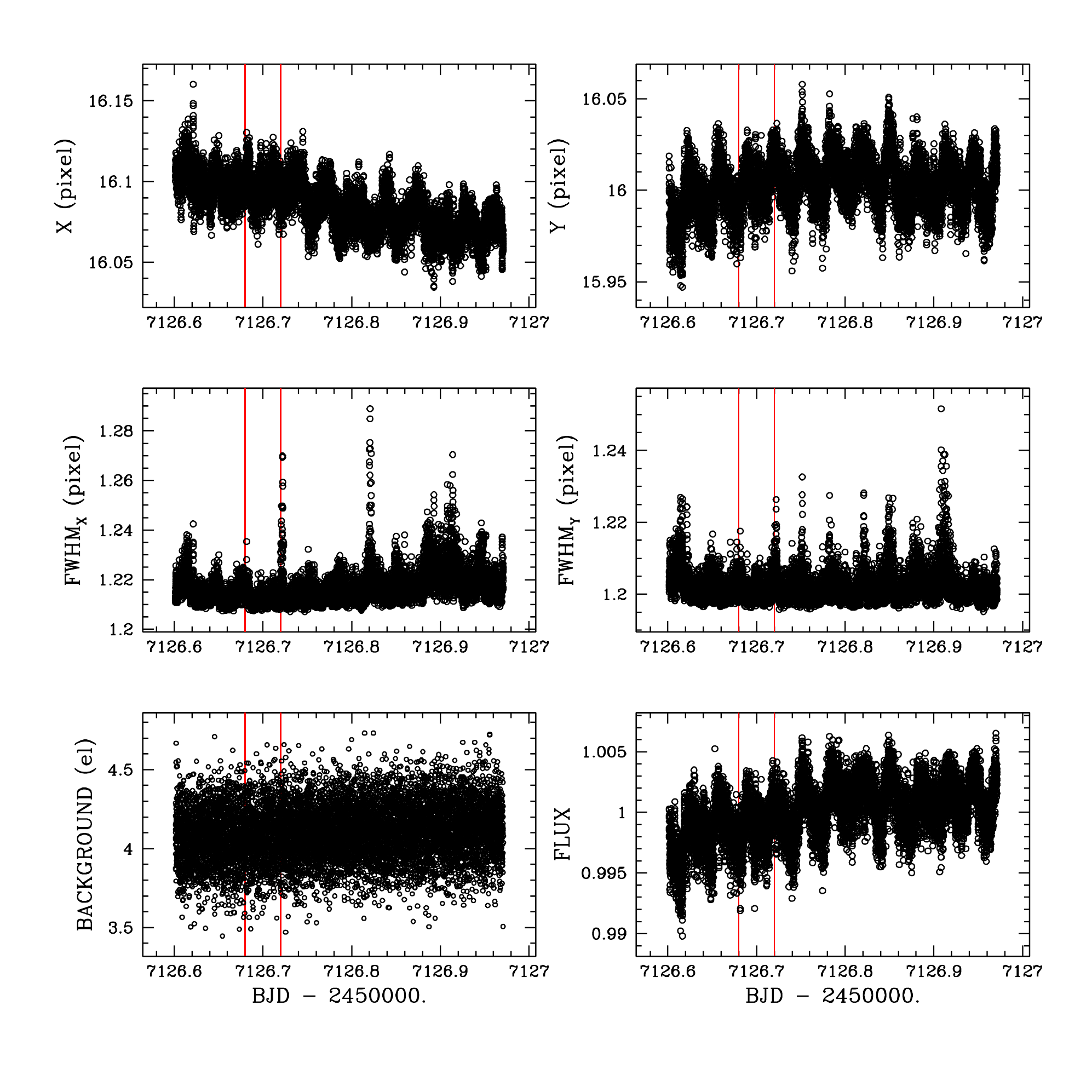}
\caption{Evolution of the following measured parameters in the {\it Spitzer} images for HD\,219134: PSF centre $x$- and $y$-positions ($top$), PSF $x$- and $y$-width ($middle$), background counts ({\it bottom left}), and stellar fluxes ({\it bottom right}). The times of the start and end of the detected transit are shown as red vertical lines.}\label{fig:S1}
\end{figure}

Following \cite{2014A&A...563A..21G}, the instrumental models included three types of low-order polynomials. The first one had as variables the $x$- and $y$-positions of the centre of the PSF to represent the `pixel phase' effect of the IRAC  InSb arrays \citep[e.g.][]{2008ApJ...673..526K}. The second one had as variables the PSF widths in the $x$- and/or the $y$-direction, its inclusion in the baseline model strongly increasing the quality of the fit for {\it Warm Spitzer} photometry \citep[see also][]{2014A&A...572A..73L}. The third, optional, function was a polynomial of the logarithm of time + a slope to represent a sharp decrease of the detector response at the start of the AOR \citep[`ramp' effect,][]{2008ApJ...673..526K}. To improve the quality of the modelling of the pixel phase effect, especially the fitting of its highest frequency components, we supplemented the $x$- and $y$-polynomial with the Bi-Linearly-Interpolated Sub-pixel Sensitivity (BLISS) mapping method \citep{2012ApJ...754..136S}. The sampling of the positions space was selected so that at least five measurements fall within the same sub-pixel. We refer the reader to \cite{2014A&A...563A..21G} for more details.  
  
\begin{figure}[t!]
\centering                     
\includegraphics[width=0.45\textwidth]{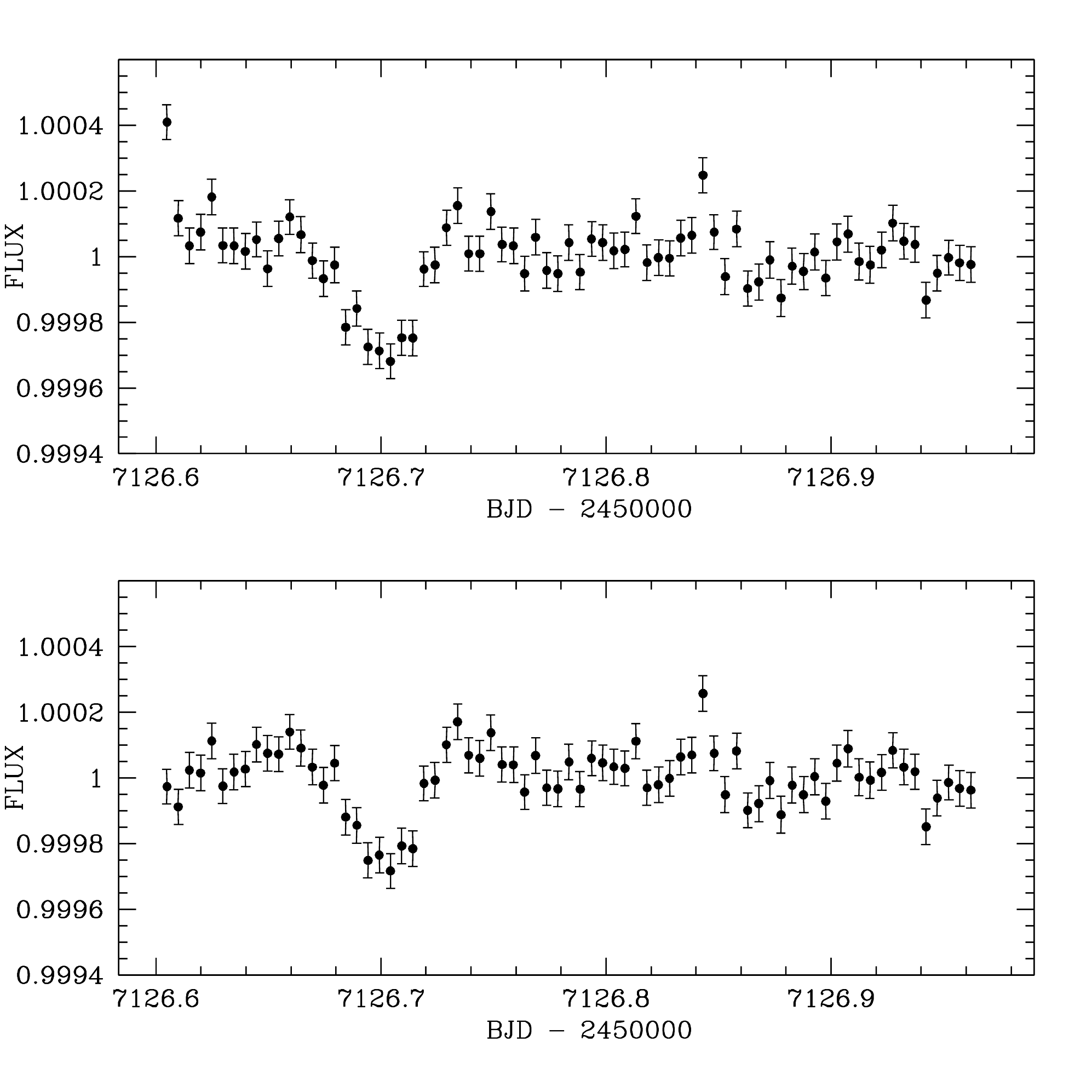}
\caption{{\it Spitzer} photometry divided by two different best-fit baseline models. The first one ($top$) aims to remove position- and PSF effects, and consists in a 4th order polynomial of the PSF $x$-and $y-$centres and widths, supplemented with the BLISS mapping method \citep{2012ApJ...754..136S} to remove high-frequency position effects. The second baseline model ($bottom$) adds to the first one a slope, and a quadratic function of the logarithm of time to model the sharp decrease of the counts at the beginning of the run (negative ramp). Both light curves are binned per 0.005d = 7.2min intervals.}\label{fig:S2}
\end{figure}

\begin{figure}
\centering      
\includegraphics[width=0.45\textwidth]{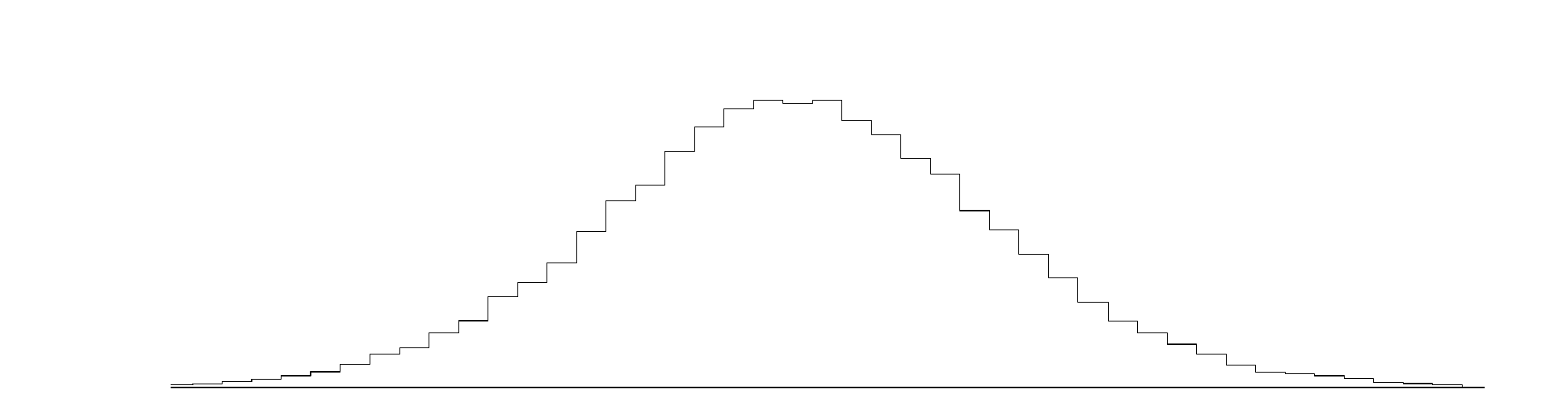}               
\includegraphics[width=0.45\textwidth]{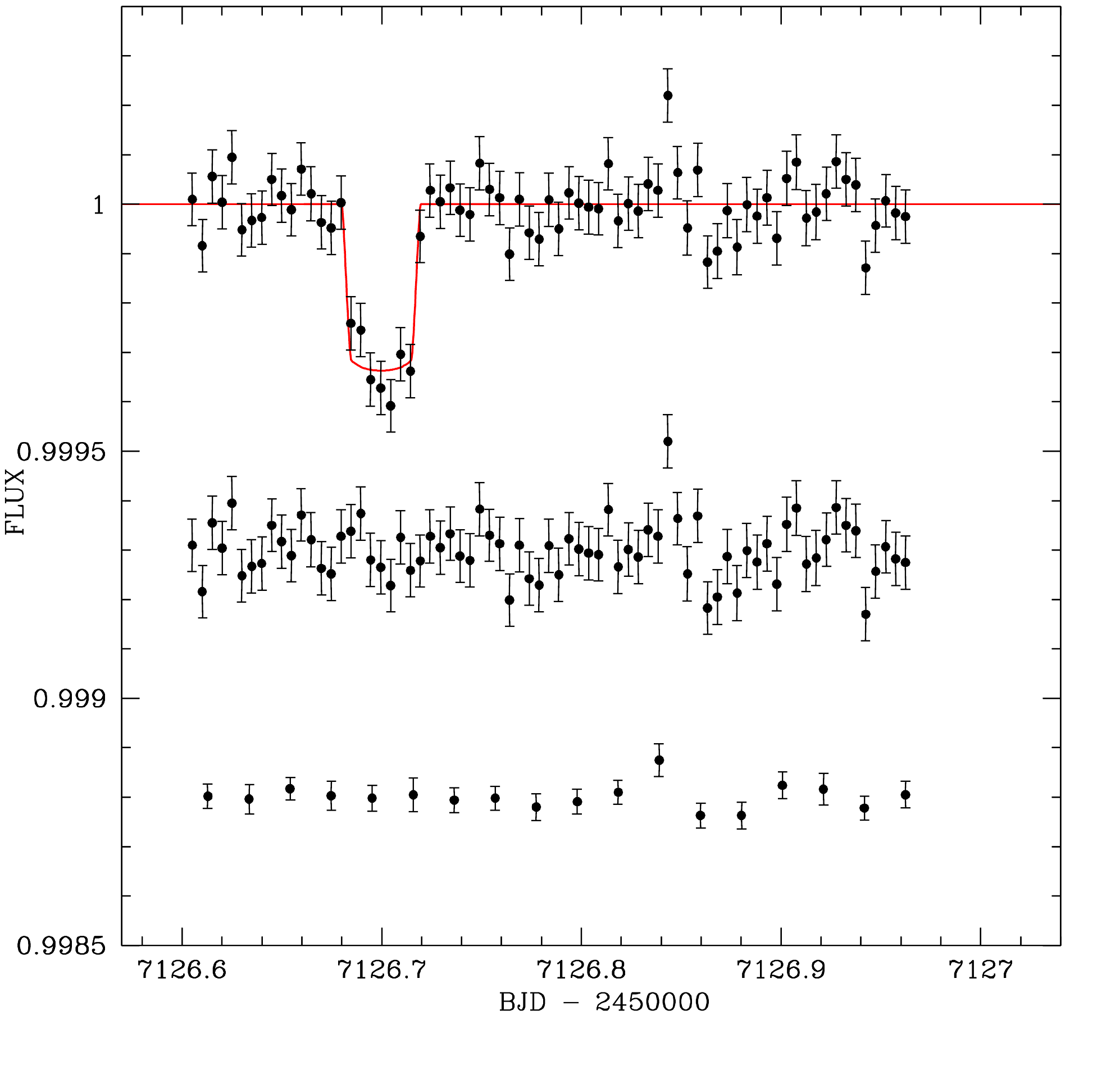}
\caption{{\it Spitzer} photometry divided by the best-fit baseline model and binned per 0.005d = 7.2min, with the best-fit transit model over-imposed in red. Below are shown the $y$-shifted residuals of the fit binned per 7.2 min and 30 min intervals. Their standard deviations are, respectively, 57 ppm and 25 ppm. Over the light curve is shown the prior probability distribution function derived for the transit timing of HD\,219134\,b from our analysis of the HARPS-N RVs.}\label{fig:S3}
\end{figure}

 Assuming no transit of HD\,219134\,b, the light curve corrected for the instrumental model described above showed a transit-like structure of $\sim300$ ppm depth and  lasting $\sim$50 min (see Fig.\,\ref{fig:S2}), this structure remaining if higher order terms are used in the polynomial functions.  Our first reflex was to check that this transit-like structure did not correspond to any odd behaviour of the external parameters, which was not the case (Fig.\,\ref{fig:S1}).  Before identifying the structure to the searched transit, we nevertheless performed a large set of short MCMC analyses assuming different baseline models, and assuming or not that the structure was a transit of the planet. For each baseline model, we computed the BIC difference between the best-fit models with and without transit to estimate the Bayes factor in favour of the transit hypothesis,  multiplying the likelihood term $e^{-0.5 \Delta BIC}$ by 9/91, the prior transit/no-transit probability ratio. In these tests, we multiplied the photometric errors by a Correction Factor (CF, see below) of 2.72, the highest value that we ever encountered in our past  experiences of high-precision photometry with  {\it Warm Spitzer}. It corresponds to an AOR targeting 55 Cnc (Demory et al., in prep.) for which the PCRS peak-up mode did not work properly. Doing so, we thus assumed that the HD\,219134 light curve was strongly affected by some correlated noise poorly reproduced by our instrumental model. At the end, the lowest value that we obtained for the Bayes factor under this extreme hypothesis was $\sim$1000 in favour of the transit hypothesis. We could thus conclude with certainty the transiting nature of the planet. 

We then performed a longer MCMC analysis to probe the posterior probability distribution of the transit parameters. The orbit of HD\,219134\,b was assumed to be circular in this MCMC analysis. The jump parameters of the MCMC, i.e. the parameters randomly perturbed at each step of the Markov Chains, were the following.
 \begin{itemize}
\item The stellar mass $M_\ast$, radius $R_\ast$, effective temperature $T_{\rm eff}$, and metallicity [Fe/H]. For these four parameters, normal prior probability distribution functions based on the values given in Table \ref{tab:star_summary} were assumed.
 \item The planet/star area ratio $dF = (R_p/R_\ast)^2$. 
 \item The impact parameter $b = a \cos{i}/R_\ast$ of the planet, where $a$ is the orbital semi-major axis and $i$ is the orbital inclination. A full-transit configuration corresponds to $b < 1 - R_p/R_\ast$.
 \item The time of inferior conjunction $T_0$ for the planet, corresponding to the mid-time of the transit. 
\end{itemize}
 The limb-darkening of the star was modeled by a quadratic law \citep{2000A&A...363.1081C}. Values for the two quadratic limb-darkening coefficients $u_1$ and $u_2$ were drawn at all steps of the MCMC from normal distributions with expectations and standard deviations  drawn from the tables of \cite{2011A&A...529A..75C} for the {\it Spitzer} 4.5 $\mu$m bandpass and for the stellar atmospheric parameters drawn at the same step. 

Five chains of 100,000 steps were performed for each analysis, their convergence being checked using the statistical test of \cite{Gelman1992}. They followed a preliminary chain of 100,000 steps performed to estimate the need to rescale the photometric errors, at the end of which the standard deviation of the residuals was compared to the mean photometric errors, and the resulting ratios $\beta_w$ were stored. $\beta_w$ represents the under-  or overestimation of the white noise of each measurement. On its side, the red noise present in the light curve (i.e. the inability of our model to represent perfectly the data) was taken into account as described in \cite{2010A&A...518A..25G}, i.e. a scaling factor $\beta_r$ was determined from the standard deviations of the binned and unbind residuals for different binning intervals ranging from 5 to 120 minutes, the largest values being kept as $\beta_r$. At the end, the error bars  were multiplied by the correction factor $CF = \beta_r \times \beta_w$. The derived values for $\beta_r $ and  $\beta_w$ were 1.30 and 1.01, resulting in $CF = 1.32$.
 
Table\,\ref{tab:transit} presents the resulting values plus error bars for the transit and planet's parameters, while Fig \ref{fig:S3} shows the light-curve corrected for the systematics, the best-fit transit model, and the residuals. 

\begin{table}
\begin{center}
\begin{tabular}{cc} \hline  
{\it Transit parameters}  & Value \\ \hline  \noalign {\smallskip} 
Depth $dF$                       &  $359 \pm 38$  ppm                                       \\ \noalign {\smallskip} 
Impact parameter $b$       &  $0.920 \pm 0.010$ $R_*$                             \\ \noalign {\smallskip} 
Timing $T_0$                    &  $2457126.7001 \pm 0.0010$ BJD$_{TDB}$ \\ \noalign {\smallskip}
Duration  $W$                   & $57.4 \pm 2.4$ min                                         \\ \hline  \noalign {\smallskip} 
{\it Physical parameters }  & \\ \hline  \noalign {\smallskip} 
Planet radius $R_p$          & $1.606 \pm 0.086$    $R_{\rm \oplus}$           \\ \noalign {\smallskip} 
Orbital inclination $i$         &  $85.058 \pm 0.080$ deg                                \\ \noalign {\smallskip} 
\hline
\end{tabular}
\caption{Transit and physical parameters deduced from the MCMC analysis of the {\it Spitzer} photometry.}
\label{tab:transit}
\end{center}
\end{table}

\section{Discussion}
\label{sec:discussion}

\subsection{Dynamical stability}
%\label{sec:dynamics}
A very important and necessary a posteriori consistency check of a planetary system characterization is needed to verify that the dynamical evolution of the system is viable on the long term, assuring the persistence of the system from the end stage of its formation (when the protoplanetary disk disappears) till its observation today. Pure n-body integrations of the 4-planet system, using both the GENGA Code \citep{2014ApJ...796...23G} launched through the DACE interface
\footnote{DACE is a platform of the National Centre for Competence in Research `PlanetS', that can be accessed at http:$\backslash$$\backslash$dace.unige.ch}, and a fourth-order Hermite scheme \citep{1991ApJ...369..200M}, with initial conditions derived from parameters in Table\,\ref{tab:planet-solution}, and assuming coplanarity and an inclination of 5 degrees from the transit observation, shows that the system is stable for more than $10^6$ orbits of the outermost planet.

A more complete analysis, taking into account general relativity and tides as well as longer-term secular effects, is beyond the scope of this paper.  As expected, a preliminary check indicates, however, that general relativity and tides will act to limit the secular growth of the eccentricities of the inner planets, favouring the long-term stability of the system.

\subsection{Bulk composition of the planet HD \,219134\,b}

\begin{figure}[t!]
\begin{center}
\includegraphics[width=0.56\textwidth]{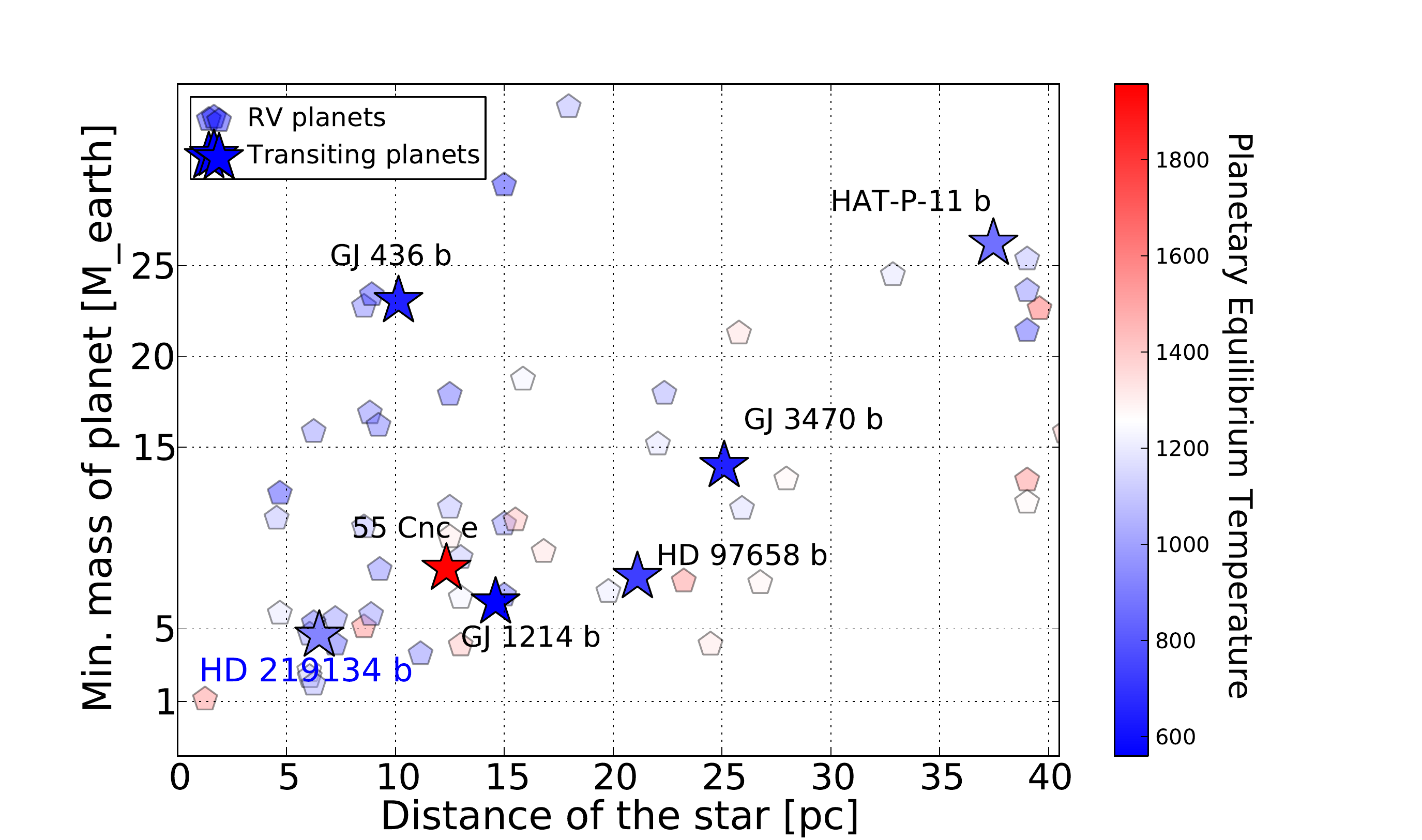}
\caption{Planet mass vs distance to the host stars for planets in close neighborhood. Transiting planets
with good mass and radius (density) determinations are shown as  $\star$ in this diagram. The planets detected by radial velocity are also shown using the minimum-mass as a proxy for the mass (data taken from www.exoplanets.org). }
\label{fig:rv_exo}
\end{center}
\end{figure}

HD \,219134\,b is the nearest transiting super-Earth known today. The radius and mass of the planet have been characterized to 6\% and 9\% accuracy, allowing us to place tight constraints on the bulk density. It thus will become one of the best targets for internal structure and atmosphere characterization with the \textit{Spitzer} and HST and the up-coming space follow-up missions e.g. TESS, JWST, CHEOPS and PLATO. Its position in a stellar distance vs planetary mass diagram is provided in Fig.\,\ref{fig:rv_exo}, in comparison with the other most favourable known cases for characterization within 40\,pc from the sun and up to 35 M$_{\oplus}$: GJ\,436\,b  \citep{Butler2004, Gillon2007}, 55\,Cnc\,e \citep{Gillon2012_55cnce,De_Mooij2014}, GJ\,1214\,b \citep{2009Natur.462..891C},  HD\,97658\,b \citep{Howard2011,Dragomir2012,Van_Grootel2014}, GJ\,3470\,b \citep{Bonfils2012, Demory2013} and HAT-P-11\,b \citep{2009ApJ...699L..48D}. 
%HD\,189733\,b \citep{2005A&A...444L..15B}, 

\begin{figure*}[t!]
\begin{center}
\includegraphics[width=0.75\textwidth]{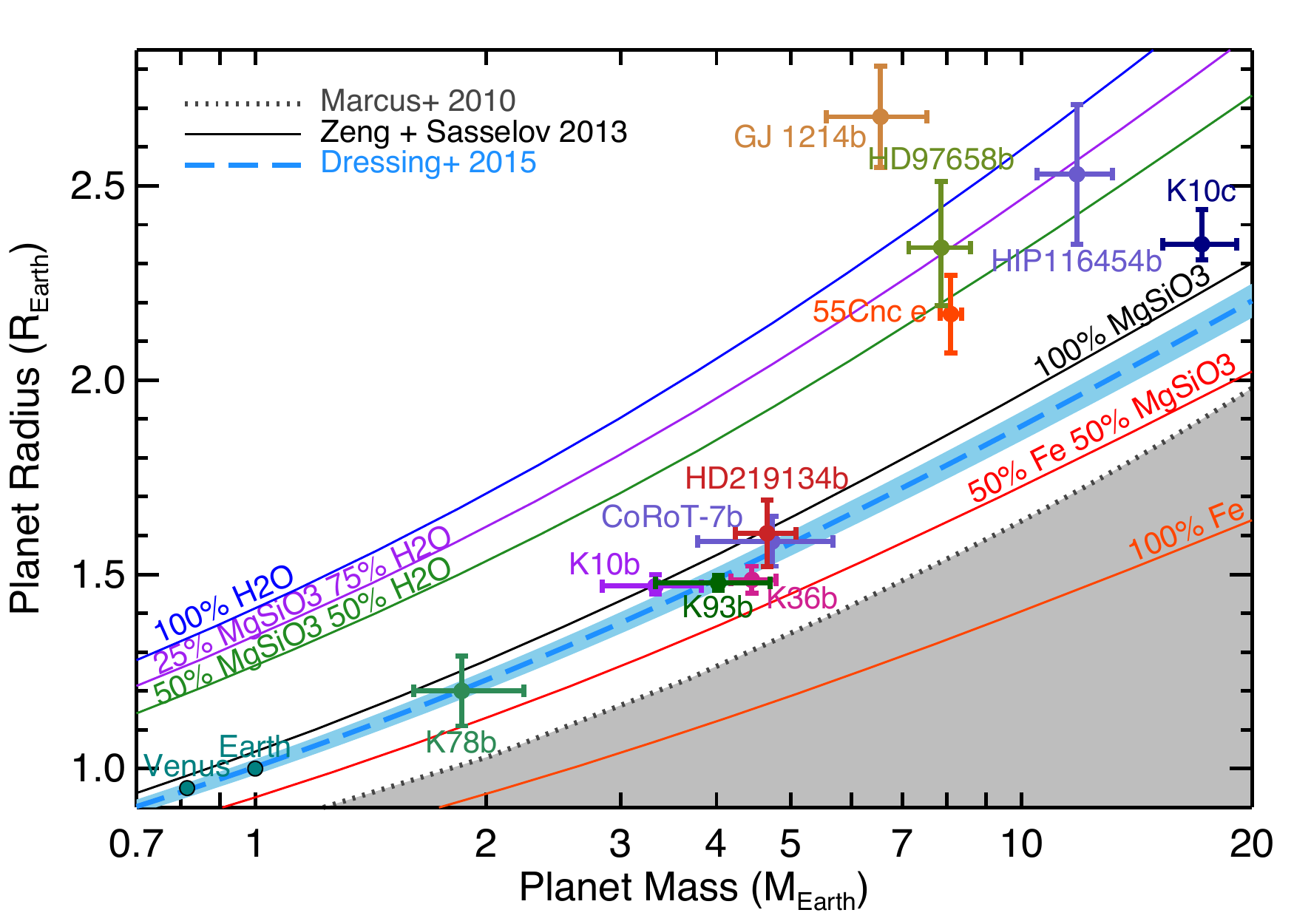}
\caption{Mass-radius relation for planets with radii smaller than 2.7\,R$_{\oplus}$ and with masses determined to a precision better than 20\,\% (updated from \cite{Dressing2015}). The shaded gray region in the lower right indicates planets with iron content exceeding the maximum value predicted from models of collisional stripping \citep{2010ApJ...719L..45M} . The solid lines are theoretical mass-radius curves \citep{2013PASP..125..227Z}  for planets with compositions of 100\,\% H$_{2}$O (blue), 25\,\% MgSiO$_3$ - 75\,\% H$_{2}$O (purple), 50\,\% MgSiO$_3$ - 50\,\% H$_{2}$O (green), 100\,\% MgSiO$_3$ (black), 50\,\% MgSiO$_3$ - 50\,\% Fe (red), and 100\,\% Fe (orange). In this diagram, the position of HD\,219134\,b is almost overlapping the point for CoRoT-7\,b. It belongs to a group of planets including Kepler-36\,b, Kepler-93\,b, and Kepler-10\,b.}
\label{fig:mass-radius}
\end{center}
\end{figure*}

Combining the spectroscopic and space-based photometric data, we estimate a density of $5.89\pm1.17$ gcm$^{-3}$ for the planet. This density is consistent with the value of 6.90 gcm$^{-3}$ that would be predicted for a $1.606$\,R$\oplus $ planet obeying the Earth-like compositional model presented in \citet{Dressing2015}. 

The compositional tracks employed in \cite{Dressing2015} (reproduced and updated in Fig.\,\ref{fig:mass-radius}) are based on interior structure models by \cite{2013PASP..125..227Z} that represent small planets as fully differentiated iron cores surrounded by lower density magnesium silicate mantles. These models provide a useful framework for comparing relative planet compositions, but the absolute core mass fractions are underestimated slightly because the \cite{2013PASP..125..227Z} models do not incorporate the presence of lighter elements in the core and the inclusion of water in the mantle.  Accordingly, the Earth-like compositional track presented in \citet{Dressing2015} corresponds to a model composition of 83\% MgSiO$_3$ and 17\% Fe whereas the actual core mass fraction of the Earth is closer to 30\%.
\citet{2015AAS...22540602Z} have recently updated their model framework to account for the presence of lighter elements in the core and the presence of water in the mantle. Employing the new models, we find that the population of highly irradiated dense planets (HD\,219134b, CoRoT-7b, Kepler-10b, Kepler-36b, Kepler-78b, and Kepler-93b) are best described by a two-component iron-magnesium silicate model with a core mass fraction of approximately 22-23\%.

\section{Conclusion}
\label{sec:conclusion}
We have presented in this paper the first result from the Rocky Planet Search (RPS) program conducted with HARPS-N, as a planetary system of three inner super-Earths and one outer sub-Saturn planet, hosted by the bright and nearby quiet K dwarf HD\,219134. The planet separations between 0.039 and 0.23\,AU called for a search of potential transits of the inner planet(s) with the \textit{Spitzer} space telescope. The successful detection of the transit of HD\,219134\,b makes the star the closest (6.5\,pc) and brightest (V=5.5) star known to date with a transiting planet (super-Earth). This system is thus becoming one of the most favourable ones for follow-up observations, for better constraining the system architecture or  aiming at the characterization of the planet physical properties. First, it provides an exquisite constraining point in the mass-radius diagram (Fig.\,\ref{fig:mass-radius}) for the composition of the planet, found to be of terrestrial-equivalent composition with a core mass fraction of the order of 22\,\%. The quality of the measurements of the radius, mass and then mean density actually foreshadows what can be expected from the future transit missions in preparation that will target bright stars (CHEOPS, TESS, PLATO). We also know from Kepler results that multi-transiting systems of small-size planets are numerous. It is then now highly suitable to search for traces of transits of the other planets in the systems. Finally, even if a potential atmosphere around the planet is expected a priori to be tiny, the brightness of the system makes it worth trying to detect features of this atmosphere in the UV, visible and NIR, from space and from the ground, especially in preparation for future measurements with larger facilities (JWST, TMT).

\begin{acknowledgements}
The HARPS-N project was funded by the Prodex program of the Swiss Space Office (SSO), the Harvard University Origin of Life Initiative (HUOLI), the Scottish Universities Physics Alliance (SUPA), the University of Geneva, the Smithsonian Astrophysical Observatory (SAO), and the Italian National Astrophysical Institute (INAF), University of St. Andrews, Queen's University Belfast, and University of Edinburgh. The research leading to these results has received funding from the European Union Seventh Framework program (FP7/2007- 2013) under grant agreement No. 313014 (ETAEARTH). C.D. is supported by a National Science Foundation Graduate Research Fellowship. P.F. acknowledges support by Funda\c{c}ao para a Ci\^encia e a Tecnologia (FCT) through Investigador FCT contracts of reference IF/01037/2013 and POPH/FSE (EC) by FEDER funding through the program "Programa Operacional de Factores de Competitividade - COMPETE". 
This work has been carried out in the frame of the National Centre for Competence in Research `PlanetS' supported by the Swiss National Science Foundation (SNSF). S.U., C.L., D.S. and F.P.  acknowledge the financial support of the SNSF.
This work is based in part on observations made with the {\it Spitzer Space Telescope}, which is operated by the Jet Propulsion Laboratory, California Institute of Technology under a contract with NASA. Support for this work was provided by NASA. M. Gillon is Research Associate at the Belgian Scientific Research Fund (F.R.S-FNRS), and he  is extremely grateful to NASA and SSC Director for having supported his searches for RV planets with  {\it Spitzer}. 
PF further acknowledges support from Funda\c{c}\~ao para a Ci\^encia e a Tecnologia (FCT) in the form of an exploratory project of reference IF/01037/2013CP1191/CT0001.
RDH was supported by STFC studentship grant ST/J500744/1 during the course of this work.
CAW acknowledges support from STFC grant ST/L000709/1.
This publication was made possible by a grant from the John Templeton Foundation. The opinions expressed in this publication are those of the authors and do not necessarily reflect the views of the John Templeton Foundation. This material is based upon work supported by the National Aeronautics and Space Administration under Grant No. NNX15AC90G issued through the Exoplanets Research Program.
\end{acknowledgements}

\bibliographystyle{aa}
\bibliography{HD219134}
 
\begin{noteaddname}
During the refereeing process, we learned about an independent detection by Vogt et al. (Laughlin, private communication) reporting additional planets in the system, based on long-term radial velocities obtained with the Keck and APF telescopes. 
\end{noteaddname}

\end{document}